\documentclass[12pt,english]{iopart}

\usepackage{iopams}
\usepackage{setstack}
\usepackage{subeqn}
\usepackage{amstext}
\usepackage{babel}
\usepackage[pdftex]{graphicx}
\usepackage{hyperref}
\usepackage{color}
\usepackage{latexsym}
\usepackage{bm}

\begin{document}

\title{Scattering by linear defects in graphene: a tight-binding approach}


\author{J. N. B. Rodrigues$^{1}$, N. M. R. Peres$^{2}$, and J. M. B. Lopes
dos Santos$^{1}$}

\address{$^{1}$ CFP and Departamento de F\'{i}sica e Astronomia, Faculdade de Ci\^{e}ncias
Universidade do Porto, P-4169-007 Porto, Portugal}

\address{$^{2}$ Physics Department and CFUM, University of Minho, P-4710-057,
Braga, Portugal}

\ead{peres@fisica.uminho.pt}

\begin{abstract}
  We develop an analytical scattering formalism for computing the
  transmittance through periodic defect lines within the tight-binding
  model of graphene.  We first illustrate the method with a relatively
  simple case, the \textit{pentagon only} defect line. Afterwards,
  more complex defect lines are treated, namely the $zz(558)$ and the
  $zz(5757)$ ones. The formalism developed, only uses simple
  tight-binding concepts, reducing the problem to matrix manipulations
  which can be easily worked out by any computational algebraic
  calculator.
\end{abstract}


\section{Introduction}

Grain boundaries (GBs) in artificially grown solids are, most likely,
unavoidable. This is particularly true for solids grown by chemical
vapor deposition. In this method, the crystal starts growing simultaneously
at different locations on the substrate. The relative orientations
of the domains have a stochastic distribution and when two domains
growing at different locations approach each other they form a GB.

If the grown crystal is used in a device larger than the size of the
grains, then an electron has to pass through one or several GBs as
it makes its way across the device. Therefore, the scattering problem
of an electron off a GB becomes technologically relevant.

In low dimensional systems, such as graphene \cite{RMP_Peres}, the
study of GBs is an active field of research \cite{NatureGrain,APLGrain}
and it has been shown that this type of disorder has a strong impact
in the transport properties of graphene \cite{NatureTransp,SSCommGrain,Louie}.
Indeed, one-dimensional defect lines can give rise to exotic effects,
such as a valley filter \cite{PhysLettADefect,PRLDefect}, where states
from one Dirac cone are filtered from those belonging to the other
inequivalent cone. It has been suggested that GBs can be exploited
for novel graphene-based nanomaterials and functional devices \cite{Louie}.

The study of scattering by extended defects is revealing itself of
increasing interest, specially after the recent work \cite{Tsen_Science:2012}
by Tsen \textit{et al.}. In the latter, using chemical vapor deposited
polycrystalline graphene, the authors made electric measurements across
a single GB. In this way, they were able to measure the electronic
properties of single GBs. Tsen \textit{et al.} have found that the
transport properties of these systems are strongly dependent on their
microscopic details. Each CVD synthesis method, typically gives rise
to GBs with similar resistivity profiles. But the GBs originating
from different synthesis procedures are normally very distinct. All
this highlights the importance of the studies presented on this paper
concerning the electronic properties of defect lines in graphene.

The study of one localized impurity in a one-dimensional system has
a simple analytical solution \cite{Sautet}, which was later generalized
to the ladder case \cite{Mizes}. This method is, however, inappropriate
for tackling more complex systems. For problems of the same nature
as the ones considered in this work, the method of non-equilibrium
Green's functions is often employed \cite{PhysLettADefect,PeresEJPB}.
Unfortunately, such method is not so easy to follow by the non-specialist,
although there is a one-to-one correspondence between the Green's
function method and the mode matching one \cite{Khomyakov}. A more
elementary method, based on the work of Ando \cite{Ando}, was applied
to the solution of a defect on ultra-narrow graphene \cite{Peres_Narrow}.
Here we develop an approach, with close resemblance to the method
of Ando for quantum point contacts, which is particularly suitable
to deal with extended periodic defects. Its starting point is the
reduction of the two-dimensional scattering problem to an effective
quasi-one-dimensional one, whose solution is simpler to work out.

In the context of the first-neighbor tight-binding model of graphene,
we study the electronic scattering from periodic defect line in graphene,
developing a systematic procedure to approach such problems. We illustrate
the method with three types of defect lines oriented along the zigzag
direction. In a companion paper we have focused on the continuum low-energy
limit of the electronic scattering from the defect lines under study
in the present text \cite{Rodrigues_PRB:2012}.

In Section \ref{sec:GenDescrp} we start by describing briefly and
in general terms how these kind of problems can be tackled. After
that introduction, we concentrate on the study of electron scattering
from a simple model of a defect line on graphene, namely the \textit{pentagon-only}
defect line (Section \ref{sec:doublepentagonGB}). We then analyze
two more complex defect lines, namely the $zz(5757)$ \cite{Rodrigues_PRB:2012}
and the $zz(558)$ ones \cite{Lahiri_NatureNanotech:2010,PRLDefect,PhysLettADefect,Rodrigues_PRB:2012}
(see Section \ref{sec:zz5757-zz558}). Finally, in Section \ref{sec:conclusion}
we summarize the results obtained and pinpoint the strengths of the
method presented in this paper.

\section{General formulation of the problem}

\label{sec:GenDescrp}

Electronic scattering from periodic extended defects in the center
of a pristine $2$D crystal is essentially equivalent to the electron
scattering from a localized defect at the center of a quasi-$1$D
crystal. Such a statement arises from the fact that the periodic extended
defect preserves crystal's translation invariance along the defect
direction. Therefore, we can Fourier transform the $2$D crystal along
the latter direction thus converting the $2$D problem into a quasi-$1$D
equivalent one.

\begin{figure}[htp!]
  \centering
  \includegraphics[width=0.85\textwidth]{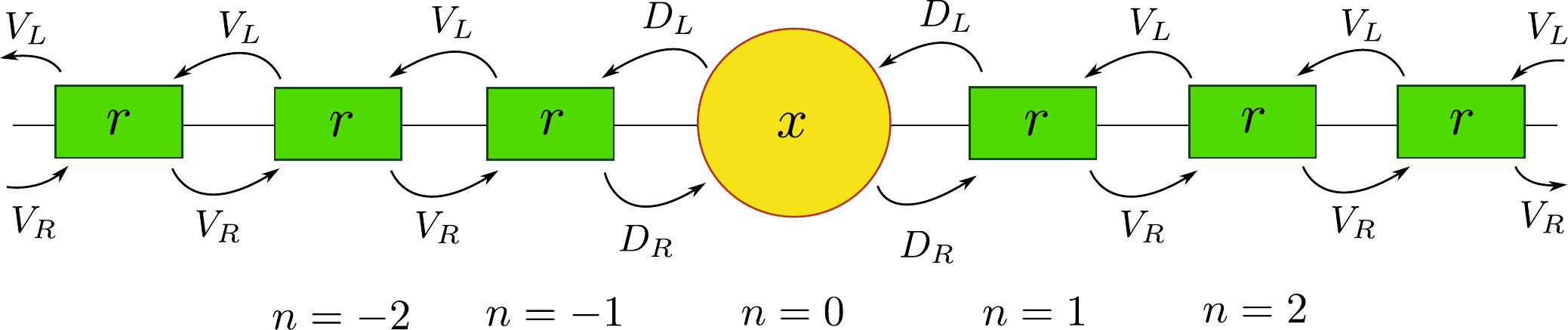}
  \caption{(Color online) Scheme of a general quasi-$1$D crystal with
    a defect at position $n=0$. The unit cell in the bulk (green
    rectangles), encompasses $r$ Wannier states, while in the defect (yellow
    circumference) there are $x$ states. The generalized hopping
    amplitude $V_{L}$ ($V_{R}=V_{L}^{\dagger}$), is a matrix that
    contains all the hopping amplitudes between Wannier states of neighbor bulk
    unit cells. The generalized hopping amplitude $D_{L}$ ($D_{R}$),
    is a $x\times r$ matrix containing the hopping amplitudes
    connecting all the Wannier states in the defect and those in the unit cell
    at $n=-1$ (n=1).}
  \label{fig:Q1Dcrystall}
\end{figure}

Thereupon, this class of $2$D problems can be treated under the framework
of scattering $1$D problems. Accordingly, in the present section
we describe, with some generality, how one can work out electron scattering
from a defect located at the center of a quasi-$1$D crystal modeled
by a first neighbor tight-binding (TB) model (see Fig. \ref{fig:Q1Dcrystall}).
This will be the starting point for the treatment of some $2$D scattering
problems in graphene (see Sections \ref{sec:doublepentagonGB} and
\ref{sec:zz5757-zz558}).

We start by writing the TB bulk equations of the general quasi-$1$D
crystal with a defect at its center (see Fig. \ref{fig:Q1Dcrystall}).
Away from the defect, these can be cast in the form 
\begin{eqnarray}
V_{R}\mathbf{c}(n-1)+(H-\epsilon\mathbb{I}_{r})\mathbf{c}(n)+V_{L}\mathbf{c}(n+1) & = & 0.\label{eq:GenTBeqs}
\end{eqnarray}
 In Eq. (\ref{eq:GenTBeqs}), $\mathbf{c}$ stands for a column vector
with as many entries as there are Wannier states in the quasi-$1$D crystal's
unit cell; we denote this number by $r$. Therefore, the terms $H$,
$\mathbb{I}_{r}$, $V_{L}$ and $V_{R}$ are $r\times r$ matrices,
$\mathbb{I}_{r}$ standing for the unit matrix. The matrix $H$ describes
the hopping processes occurring inside the unit cell, while $V_{L}$
($V_{R}$) describes the hopping processes occurring between the unit
cell at position $n$ and the unit cell at position $n+1$ ($n-1$).
The hermitian nature of the Hamiltonian requires that $V_{R}=V_{L}^{\dagger}$.

\begin{figure}[htp!]
  \centering
  \includegraphics[width=0.85\textwidth]{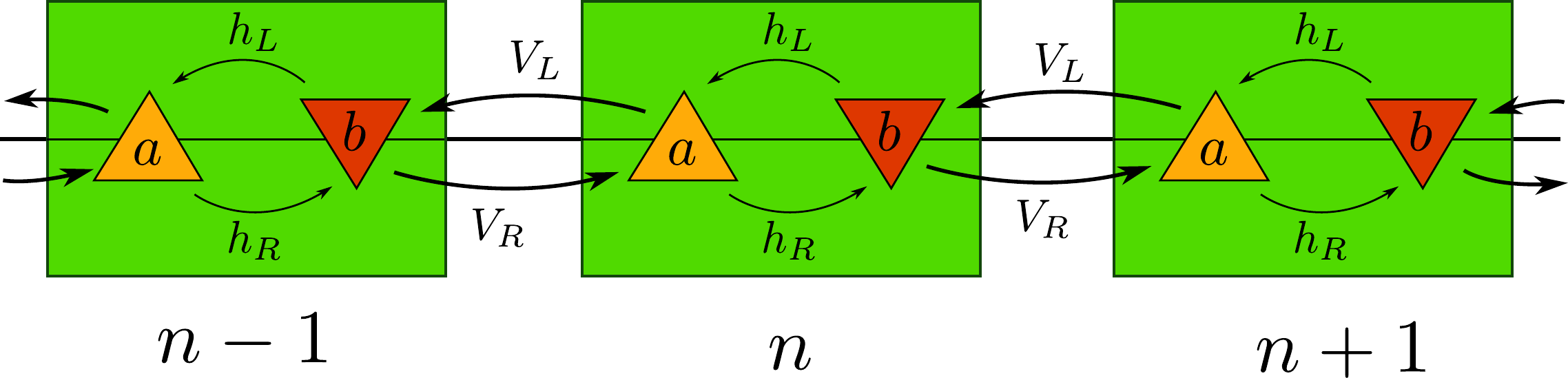}
  \caption{(Color online) Scheme of a general quasi-$1$D $AB$
    crystal. The unit cell of the crystal (green rectangles)
    encompasses $r=a+b$ Wannier states.  The generalized hopping amplitude
    $V_{L}$ ($V_{R}=V_{L}^{\dagger}$), is a matrix that contains all
    the hopping amplitudes between Wannier states of neighboring unit
    cells. These generalized hopping amplitudes, are only non-zero
    when connecting the $a$ and $b$ Wannier states of different unit cells. In
    contrast, they are zero both between the $a$ Wannier states of two
    neighboring unit cells, and between $b$ Wannier states of neighboring unit
    cells.}
  \label{fig:Q1Dcrystal-AB}
\end{figure}

The TB bulk equation {[}Eqs. (\ref{eq:GenTBeqs}){]} can usually be
written in a different form, where the amplitudes of the unit cell
located at position $n+1$ are expressed in terms of the amplitudes
of the unit cells located at positions $n$ and $n-1$, 
\begin{eqnarray}
\mathbf{c}(n+1) & = & \mathbb{T}_{1}\mathbf{c}(n)+\mathbb{T}_{2}\mathbf{c}(n-1),\label{eq:TransfMatGen}
\end{eqnarray}
 where $\mathbb{T}_{i}$ are $r\times r$ matrices, which we call
transfer matrices. These matrices are generally non-hermitian. Usually
Eq. (\ref{eq:TransfMatGen}) cannot be obtained directly from Eq.
(\ref{eq:GenTBeqs}) for site $n$, because the matrix $V_{L}$ is
not invertible. However, for the cases of interest, we can in general
obtain Eq. (\ref{eq:TransfMatGen}) from the TB equations {[}Eq. (\ref{eq:GenTBeqs}){]}
for the sites $n$ and $n+1$. When the one-dimensional chain is $AB$-like
(see Fig. \ref{fig:Q1Dcrystal-AB}) the transfer matrix description
further simplifies to 
\begin{eqnarray}
\mathbf{c}(n+1) & = & \mathbb{T}\mathbf{c}(n),\label{eq:TransfMat-AB}
\end{eqnarray}
where there is now only one transfer matrix $\mathbb{T}$ relating
neighbor amplitudes. Since in the present text we are specially interested
in studying graphene crystals, whose Fourier transformed systems give
rise to $AB$-like chains, we will from now on simplify our analysis
by assuming that our general $1$D chain is described by a relation
of the form of Eq. (\ref{eq:TransfMat-AB}).

Inherent to the study of any scattering problem is the determination
of the transverse propagating (or evanescent) modes allowed in the
system. In the case of a one-dimensional periodic chain the computation
of the modes can be done using Bloch's theorem. In that prescription,
we write the amplitudes at position $n+1$ and $n-1$ in terms of
those at position $n$: $\mathbf{c}(n+1)=e^{ik_{j}a}\mathbf{c}(n)\equiv\lambda_{j}\mathbf{c}(n)$
and $\mathbf{c}(n-1)=e^{-ik_{j}a}\mathbf{c}(n)\equiv\mathbf{c}(n)/\lambda_{j}$,
where $k_{j}$ stands for the wave-number along the chain direction
associated with the mode indexed by $j$, and 
$a$ stands for the length of the primitive vector. Using these relations, we
can rewrite Eq. (\ref{eq:GenTBeqs}) as 
\begin{eqnarray}
\frac{1}{\lambda_{j}}V_{R}\mathbf{c}(n)+(H-\epsilon\mathbb{I}_{r})\mathbf{c}(n)+\lambda_{j}V_{L}\mathbf{c}(n) & = & 0.\label{eq:GenTBeqs-lambda}
\end{eqnarray}
The $r$ eigenvalues $\lambda_{j}=e^{ik_{j}a}$ 
and their $r$ associated
eigenvectors $\vert\psi_{j}\rangle$ permitted at a given energy $\epsilon$,
are obtained by solving this generalized eigenproblem. On the other
hand, when one is able to write the TB equations in the form of a
recurrence relation of the form of Eq.~(\ref{eq:TransfMat-AB}),
it becomes obvious that the $\lambda_{j}$ and $\vert\psi_{j}\rangle$
are, respectively, the eigenvalues and right eigenvectors of the transfer
matrix, $\mathbb{T}$.

Let us now turn our attention to the defect, so that we can determine
the boundary condition that the modes should satisfy when scattering
from it. We start by assuming that regardless of the complexity of
the defect at the center of the $1$D crystal, the latter is located
at position $n=0$ (see Fig. \ref{fig:Q1Dcrystall}). Therefore, the
TB equations associated with the amplitudes at the defect can be generally
written as 
\begin{eqnarray}
D_{R}\mathbf{c}(-1)+(H_{D}-\mathbb{I}_{r})\mathbf{d}(0)+D_{L}\mathbf{c}(1) & = & 0.\label{eq:GenTBeqsDefect}
\end{eqnarray}
 In the above equation, $D_{L}$ ($D_{R}$) is a $x\times r$ rectangular
matrix containing all possible hopping amplitudes connecting the defect
Wannier states and those of the unit cell located at position $n=+1$ ($n=-1$).
The matrix $H_{D}$ encompasses all the hopping amplitudes between
the $x$ Wannier states of the defect, whose amplitudes are grouped in the
$x$-dimensional array $\mathbf{d}(0)$. Note that Eq. (\ref{eq:GenTBeqsDefect})
is a shorthand for a system of equations, containing $x$ TB equations
associated with the $x$ states at the defect and $x+2r$ unknown amplitudes.
As a consequence, if we want to write a passage equation relating
the amplitudes at $\mathbf{c}(1)$ and those at $\mathbf{c}(-1)$,
we will also have to use the TB equations at position $n=-1$ 
\begin{eqnarray}
V_{R}\mathbf{c}(-2)+(H-\mathbb{I}_{r})\mathbf{c}(-1)+D_{L}\mathbf{d}(0) & = & 0,\label{eq:GenTBeqsDefect-m2}
\end{eqnarray}
 as well as the transfer matrix relation $\mathbf{c}(-1)=\mathbb{T}\mathbf{c}(-2)$.
Doing so, we end up with $x+3r$ unknown amplitudes and $x+2r$ equations.
We choose to solve them by expressing $\mathbf{c}(1),\,\mathbf{d}(0)$
and $\mathbf{c}(-2)$ ($x+2r$ amplitudes) in terms of $\mathbf{c}(-1)$.
Therefore, we will be able to write a passage equation relating the
amplitudes before and after the defect as 
\begin{eqnarray}
\mathbf{c}(1) & = & \mathbb{M}\mathbf{c}(-1),\label{eq:DfctPassageRelGen}
\end{eqnarray}
where $\mathbb{M}$ is a $r\times r$ matrix. It is natural to interpret
the latter matrix as a boundary condition matrix imposed by the defect
on the wave function at each one of its sides. 

We have so far shown how one can determine both the scattering modes
of the one-dimensional chain and the boundary condition they must
obey at the defect. The only step remaining is the computation of
the scattering coefficients. Let us suppose that we have an incoming
mode from $n=-\infty$. We then expect to have both transmitted modes
outgoing to $n=+\infty$ and reflected modes outgoing to $n=-\infty$.  
As a consequence, we can write the wave function at each side of the
defect as \begin{subequations} \label{eq:WFs} 
\begin{eqnarray}
\vert\Psi(n)=\vert\Psi^{L}(n)\rangle & \equiv & \lambda_{i>}^{n+1}\vert\psi_{i}^{>}\rangle+\sum_{j=1}^{r_{<}}\rho_{ij}\lambda_{j<}^{n+1}\vert\psi_{j}^{<}\rangle,\qquad\textrm{for }n\leq-1\label{eq:WFleft}\\
\vert\Psi(n)\rangle=\vert\Psi^{R}(n)\rangle & \equiv & \sum_{j=1}^{r_{>}}\tau_{ij}\lambda_{j>}^{n-1}\vert\psi_{j}^{>}\rangle,\qquad\qquad\textrm{for }n\ge1\label{eq:WFright}
\end{eqnarray}
 \end{subequations} where the superscript $L$ stands for the wave function
to the \textit{left} of the defect ($n<0$), while the superscript
$R$ stands for the wave function to the right of the defect ($n>0$).
In Eqs. (\ref{eq:WFs}), the sum over $j$ up to $r_{>}$ ($r_{<}$)
stands for a sum over all the $p_{>}$ ($p_{<}$) propagating modes
moving in the direction of $n=+\infty$ ($n=-\infty$) and all the
$s_{>}$ ($s_{<}$) evanescent modes, decreasing (increasing) with
increasing $n$. Note that the total number of propagating and evanescent
modes must add up to $r$, the total number of modes: $r=r_{>}+r_{<}=p_{>}+s_{>}+p_{<}+s_{<}$,
which in the $AB$ chain happens to be equal to the number of Wannier states
in the unit cell. Moreover, the coefficients $\rho_{ij}$ ($\tau_{ij}$)
stand for the reflection (transmission) coefficients of the modes
indexed by $j$ for an incoming mode $i$.

Before proceeding, let us emphasize that in this text, we are going to use interchangeably
two types of notation: we will use the vectorial notation when referring to the arrays of 
amplitudes at a given unit cell located at position $n$ [see Eqs. (\ref{eq:GenTBeqs})-(
\ref{eq:DfctPassageRelGen})]; the Dirac {\it ket} notation will be used when referring to the 
eigenmodes of the one-dimensional chain, as well as to the wave function at a given position 
$n$ [see Eqs. (\ref{eq:WFs})].

As long as we know the mathematical expressions for the modes of the
quasi-one-dimensional chain, $\vert\psi_{j}^{>}\rangle$ and $\vert\psi_{j}^{<}\rangle$,
and the boundary condition matrix $\mathbb{M}$, the determination
of the scattering coefficients in Eqs. (\ref{eq:WFs}), $\rho_{ij}$
and $\tau_{ij}$, is a straightforward calculation. 

If we define a matrix $U$, with columns which are the $r$
transverse modes 
\begin{eqnarray}
U&=&\Big[\vert\psi_{1}^{>}\rangle,\ldots,\vert\psi_{r_{>}}^{>}\rangle,\vert\psi_{1}^{<}\rangle,\ldots,\vert\psi_{r_{<}}^{<}\rangle\Big],
\end{eqnarray}
and the following two vectors,
\begin{subequations}
\begin{eqnarray}
\vert\Phi^{L}(n)\rangle & = & \Big[0,\ldots0,\lambda_{i>}^{n+1},0,\ldots,0,\rho_{i1}\lambda_{1<}^{n+1},\ldots,\rho_{ir_{<}}\lambda_{r_{<}<}^{n+1}\Big]^{T},\\
\vert\Phi^{R}(n)\rangle & = & \Big[\tau_{i1}\lambda_{1>}^{n-1},\ldots,\tau_{ir_{>}}\lambda_{r_{>}>}^{n-1},0,\ldots,0\Big]^{T},
\end{eqnarray}
\end{subequations}
it is clear that
\begin{subequations}
\begin{eqnarray}
\vert\Psi^{L}(n)\rangle & \equiv & U\vert\Phi^{L}(n)\rangle,\label{eq:WF2left-1}\\
\vert\Psi^{R}(n)\rangle & \equiv & U\vert\Phi^{R}(n)\rangle.\label{eq:WF2right-1}
\end{eqnarray}
\end{subequations}
 The passage equation, \ref{eq:DfctPassageRelGen}, then becomes
\begin{eqnarray}
\vert\Phi^{R}(1)\rangle & = & U^{-1}\mathbb{M}U\vert\Phi^{L}(-1)\rangle .\label{eq:BCGen_PropBasis}
\end{eqnarray}
This is a non-homogeneous system of $r$ linear equations with $r$
unknowns, $\rho_{i1},\dots,\rho_{ir_{<}}$ and $\tau_{i1},\dots,\tau_{ir_{>}}$,
which we solve to obtain these scattering amplitudes.

When carrying out these calculations in a computer algebra system,
the following remark may be useful. Since the transfer matrix is non-hermitian,
its eigenbasis is not orthogonal. Nevertheless we can define a dual
basis $\vert\widetilde{\psi}_{j}\rangle,\, j=1,\dots,r$ by the relation
\begin{eqnarray}
\langle \widetilde{\psi}_{j}\vert\psi_{i}\rangle&=&\delta_{ij}
\end{eqnarray}
and it is clear that the matrix $U^{-1}$ $j^{\textrm{th}}$ row is just the vector
$\langle \widetilde{\psi}_{j}\vert$, seen that the $i^{\textrm{th}}$
column of $U$ is $\vert\psi_{i}\rangle$; in the \ref{app:DualBasis},
we show that these vectors $\langle \widetilde{\psi}_{j}\vert$
can be simply obtained as the right eigenvectors of the transpose
of the transfer matrix $\mathbb{T}^{T}$. 

It is worth noting that the boundary condition arising from the presence
of the defect, Eq. (\ref{eq:DfctPassageRelGen}), must conserve the
particle current (see \ref{app:ConsvCurr}). Equivalently, the current
on the left hand side of the defect, $\mathcal{J}_{L}\equiv\mathbf{\mathcal{J}}(-1)=\left\langle \psi(-1)\right|\hat{\mathbf{\mathcal{J}}}\vert\psi(-1)\rangle$,
must be equal to the current on the right hand side of the defect,
$\mathcal{J}_{R}\equiv\mathcal{J}(1)=\left\langle \psi(1)\right|\hat{\mathbf{\mathcal{J}}}\vert\psi(1)\rangle$,
where $\hat{\mathbf{\mathcal{J}}}$ stands for the current operator.
Therefore, the boundary condition matrix, $\mathbb{M}$, must satisfy
the following equality 
\begin{eqnarray}
\mathbb{M}^{\dagger}.\hat{\mathcal{J}}.\mathbb{M} & = & \hat{\mathbf{\mathcal{J}}}.\label{eq:CurrConsvBCcondt}
\end{eqnarray}

In the following sections we solve three electronic scattering problems
using the machinery just presented. We will see that it simplifies
the mathematical treatment of such problems without leading to a substantial
loss of physical insight and intuition over the physical phenomena
going on. On one hand, using the transfer matrix formalism we are
able to write the scattering modes $\lambda_{j}\vert\psi_{j}\rangle=e^{ik_{j}a}\vert\psi_{j}\rangle$
directly in terms of the energy $\epsilon$ and the wave-number along
the defect line, $k_{x}$. We thus avoid to write the former in terms
of the wave-number perpendicular to the defect line, $\mathbf{k}\cdot\mathbf{u}_{2}$.
On the other hand, the described procedure allows us to compute an
expression relating the TB amplitudes at the two sides of the defect.
This has the obvious interpretation as being the boundary condition
imposed on the wave function at the defect. As we will see in Sections
\ref{sec:doublepentagonGB} and \ref{sec:zz5757-zz558}, such a boundary
condition relation can be obtained after some simple algebraic manipulations
of the TB equations at the defect.

With this in mind, let us now apply this formalism to three types
of zigzag oriented defect lines in graphene: the \textit{pentagon-only}
defect line (see Fig. \ref{fig_Scheme_doubleJS_GB}), the $zz(5757)$
defect line (see left panel of Fig. \ref{fig:Scheme_zz5757-zz558})
and the $zz(558)$ defect line (see right panel of Fig. \ref{fig:Scheme_zz5757-zz558}).

\section{Scattering from a \textit{pentagon-only} defect line: a pedagogical
example}

\label{sec:doublepentagonGB}

In what follows, we shall illustrate the procedure described in Section
\ref{sec:GenDescrp} by working out the scattering of electrons from
a \textit{pentagon-only} defect line in the context of the first neighbor
tight-binding model of graphene (see Fig. \ref{fig_Scheme_doubleJS_GB}).
As this system is invariant under translations along the defect line,
Bloch's theorem allows the Fourier transformation along that direction.
Therefore, the $2$D graphene layer can be transformed into an effective
$1$D chain (see Fig. \ref{fig_Scheme_1Dchain_doubleJSGB}). This
effective $1$D chain depends on the quantum number $k_{x}$, the
wave-number along $\mathbf{u}_{1}$ direction. Based on this $1$D
effective description, we can straightforwardly employ the procedure
described in Section \ref{sec:GenDescrp}.

\subsection{Formulating the problem}

\label{sec:JS-form}

\begin{figure}[htp!]
 \centering \includegraphics[width=0.6\textwidth]{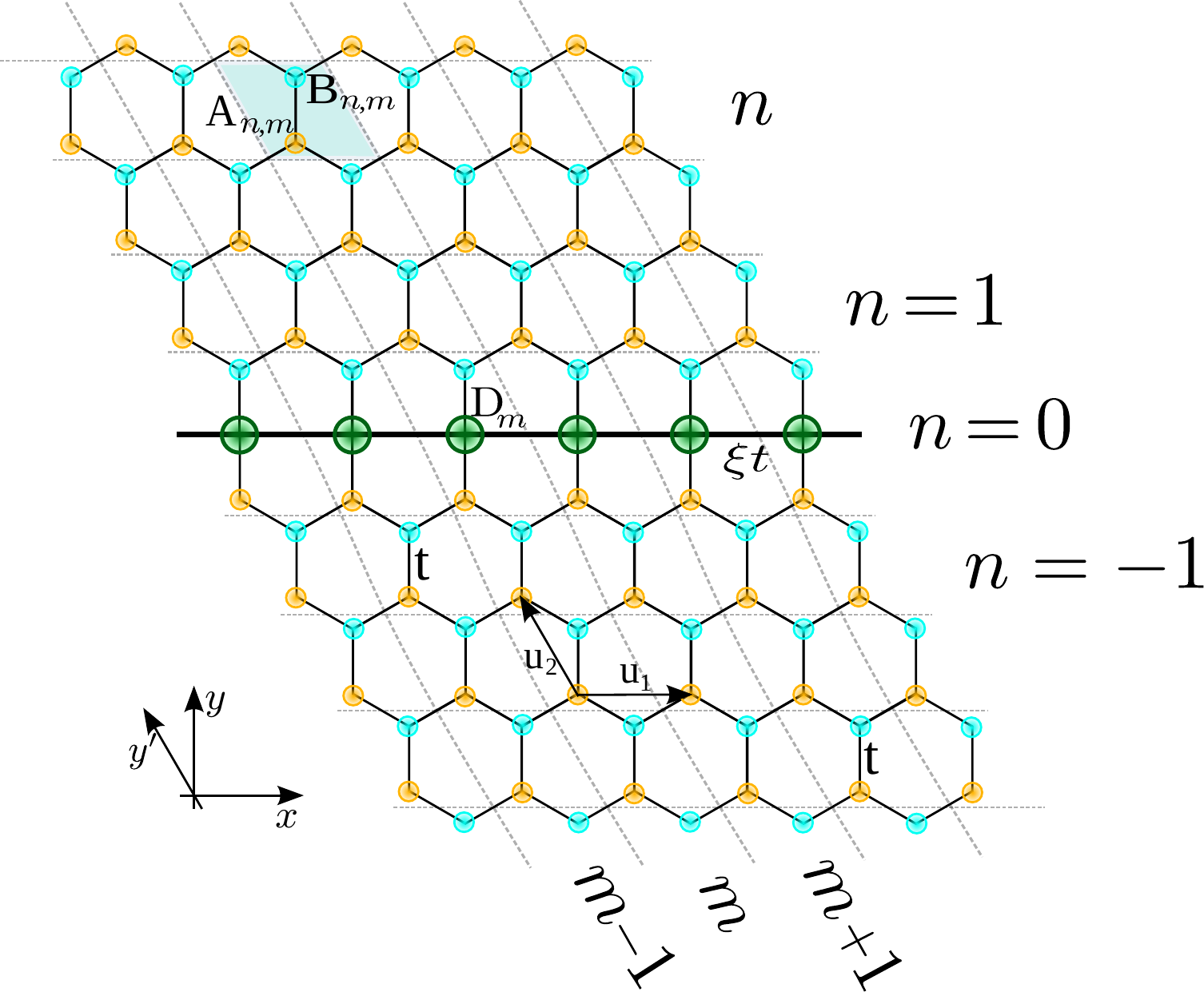}
\caption{(Color online) Graphene's honeycomb lattice with a zigzag oriented
linear defect, which we dub \textit{pentagon-only} defect line.
In the bulk the hopping is $t$, while between atoms of the defect
it is $\xi t$. The lattice vectors are denoted by $\mathbf{u}_{1}=a(1,0)$
and $\mathbf{u}_{2}=(-1,\sqrt{3})a/2$, where $a$ stands for the
lattice parameter.}

\label{fig_Scheme_doubleJS_GB} 
\end{figure}

Let us assume we have the following system: a defect line of the type
represented in Fig. \ref{fig_Scheme_doubleJS_GB}, in an otherwise
perfect graphene lattice. 

As we can see by inspection of Fig. \ref{fig_Scheme_doubleJS_GB},
the lattice vectors $\mathbf{u}_{1}$ and $\mathbf{u}_{2}$, in cartesian
coordinates, read \begin{subequations} \label{eq:LatticeVecs} 
\begin{eqnarray}
\mathbf{u}_{1} & = & a(1,0),\\
\mathbf{u}_{2} & = & a\bigg(-\frac{1}{2},\frac{\sqrt{3}}{2}\bigg),
\end{eqnarray}
 \end{subequations} where $a$ stands for the lattice parameter.

Using this coordinate system, we can write the first neighbor tight-binding
Hamiltonian of the system sketched in Fig. \ref{fig_Scheme_doubleJS_GB}
as $\hat{H}=\hat{H}^{U}+\hat{H}^{D}+\hat{H}^{L}$, where $\hat{H}^{U}$
($\hat{H}^{L}$) stands for the Hamiltonian above (below) the defect
line, while the remaining term, $\hat{H}^{D}$, describes the defect
line itself. In second quantization the explicit forms of $\hat{H}^{U}$
and $\hat{H}^{L}$ read 
\begin{eqnarray}
\hat{H}^{U(L)} & = & -t\sum_{m}\sum_{n}\bigg\{\Big[\hat{b}^{\dagger}(m,n)+\hat{b}^{\dagger}(m,n-1)\nonumber \\
 &  & +\hat{b}^{\dagger}(m-1,n-1)\Big]\hat{a}(m,n)+h.c.\bigg\}\,,
\end{eqnarray}
 where for $H^{U}$ ($H^{L}$), $n\geq1$ ($n\leq-1$). Moreover,
$H^{D}$ reads 
\begin{eqnarray}
\hat{H}^{D} & = & -t\sum_{m}\bigg\{\Big[\xi\hat{d}^{\dagger}(m+1)+\hat{a}^{\dagger}(m,0)+\hat{b}^{\dagger}(m,0)\Big]\hat{d}(m)+h.c.\bigg\},
\end{eqnarray}
 where $t$ is the usual hopping amplitude of pristine graphene, while
$\xi t$ is the hopping amplitude between the $D_{m}$ atoms of the
defect line, as represented in Fig. \ref{fig_Scheme_doubleJS_GB}.

\begin{figure}[htp!]
 \centering \includegraphics[width=0.95\textwidth]{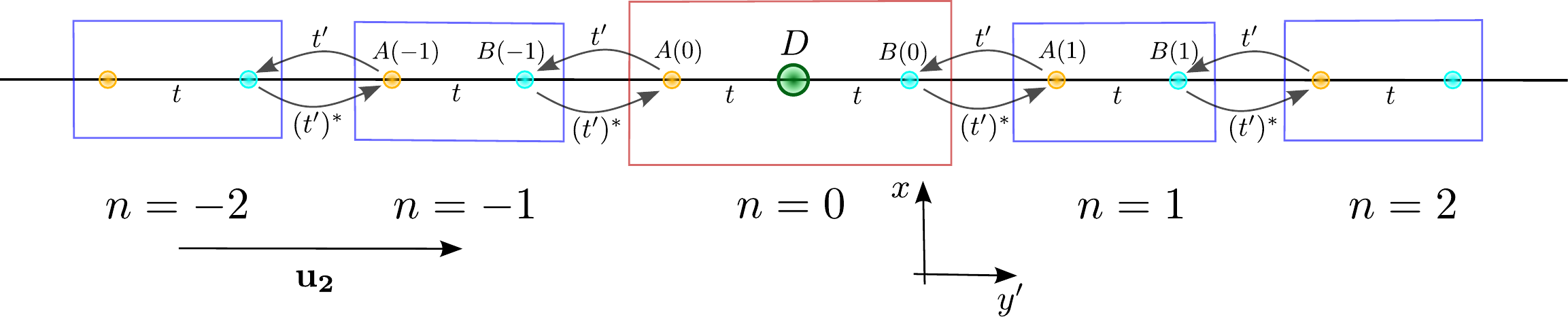}
\caption{(Color online) Effective one-dimensional chain obtained after Fourier
transforming the Hamiltonian of graphene with a \textit{pentagon-only}
defect line, along the $\mathbf{u}_{1}$ direction.}

\label{fig_Scheme_1Dchain_doubleJSGB} 
\end{figure}

As was referred in the beginning of this section, this system is invariant
under translations $\mathbf{r}=m\mathbf{u}_{1}$ (where $m$ is an
integer). Thus, we can make use of Bloch's theorem and Fourier transform
the Hamiltonian along $\mathbf{u}_{1}$. This diagonalizes the problem
relatively to the variable $m$, introducing a new quantum number,
$k_{x}$. Consequently, the resulting Hamiltonian turns out to be
equivalent to the Hamiltonian of a one-dimensional chain with two
atoms per unit cell and a localized defect at its center (see Fig.
\ref{fig_Scheme_1Dchain_doubleJSGB}).

The Hamiltonian of the effective one-dimensional chain, can be written
as $\hat{H}(k_{x})=\hat{H}^{U}(k_{x})+\hat{H}^{D}(k_{x})+\hat{H}^{L}(k_{x})$.
Its three terms read \begin{subequations} \label{eq:Hparts_1D} 
\begin{eqnarray}
\hat{H}^{U(L)}(k_{x}) & = & -\sum_{n}\bigg\{\Big[t'\hat{b}^{\dagger}(k_{x},n-1)+t\hat{b}^{\dagger}(k_{x},n)\Big]\hat{a}(k_{x},n)+h.c.\bigg\},\\
\hat{H}^{D}(k_{x}) & = & -2\xi t\cos(k_{x}a)\hat{d}^{\dagger}(k_{x})\hat{d}(k_{x})-\Big[t\hat{a}^{\dagger}(k_{x},0)\hat{d}(k_{x})\nonumber \\
 &  & +t\hat{b}^{\dagger}(k_{x},0)\hat{d}(k_{x})+h.c.\Big].
\end{eqnarray}
 \end{subequations} This one-dimensional chain has alternating hopping
amplitudes between the atoms, $t$ and $t'=t(1+e^{ik_{x}a})$. At
the defect, there is a on-site energy term, $\widetilde{\epsilon}(k_{x})=-2\xi t\cos(k_{x}a)$,
which depends on the value of the longitudinal momentum $k_{x}$.

The TB equations of such system are, as usually, obtained from Schrödinger's
equation 
\begin{eqnarray}
\hat{H}(k_{x})\vert\mu,k_{x}\rangle & = & \epsilon_{\mu,k_{x}}\vert\mu,k_{x}\rangle.\label{eq:SchrodingerEq}
\end{eqnarray}
 In Eq. (\ref{eq:SchrodingerEq}) $\hat{H}(k_{x})$ stands for the
Hamiltonian of the effective one-dimensional chain, while the eigenstate
$|\mu,k_{x}\rangle$ can be expressed as a linear combination of the
site amplitudes along the one-dimensional chain 
\begin{eqnarray}
\vert\mu,k_{x}\rangle & = & \sum_{n=-\infty}^{+\infty}\Big[A(k_{x},n)\vert a;k_{x},n\rangle+B(k_{x},n)\vert b;k_{x},n\rangle\Big]+D(k_{x})\vert d;k_{x}\rangle,
\end{eqnarray}
 where the $\vert c;k_{x},n\rangle=\hat{c}^{\dagger}(k_{x},n)\vert0\rangle$
stands for the one-particle states at the atom $c=a,b,d$ of unit
cell $n$ of the one-dimensional chain.

\subsection{Bulk properties}

\label{sec:JS-bulk}

From what we have written above, it is simple to conclude that the
TB equations in the bulk ($n\neq0$) of the one-dimensional chain
(see Fig. \ref{fig_Scheme_1Dchain_doubleJSGB}), read \begin{subequations}
\label{eq:TBbulkEqs} 
\begin{eqnarray}
\epsilon A(k_{x},n) & = & -tB(k_{x},n)-(t')^{*}B(k_{x},n-1),\label{eq:TBbulkEqs1}\\
\epsilon B(k_{x},n) & = & -tA(k_{x},n)-t'A(k_{x},n+1).\label{eq:TBbulkEqs2}
\end{eqnarray}
 \end{subequations}

To recast these equations in the form of a transfer matrix relation,
we solve Eq. (\ref{eq:TBbulkEqs2}) for $A(k_{x},n+1)$ and Eq. (\ref{eq:TBbulkEqs1})
for $B(k_{x},n)$ 
\begin{subequations} \label{eq:TBbulkEqsMod} 
\begin{eqnarray}
A(k_{x},n+1) & = & -\frac{t}{t'}A(k_{x},n)-\frac{\epsilon}{t'}B(k_{x},n).\label{eq:TBbulkEqsMod2}\\
B(k_{x},n) & = & -\frac{\epsilon}{t}A(k_{x},n)-\frac{(t')^{*}}{t}B(k_{x},n-1),\label{eq:TBbulkEqsMod1}
\end{eqnarray}
 \end{subequations} where we have used $t'=t(1+e^{ik_{x}a})$. If
now we write Eq. (\ref{eq:TBbulkEqsMod1}) for $n+1$ and substitute
Eq. (\ref{eq:TBbulkEqsMod2}) in it, we obtain a recurrence relation
between the amplitudes at unit cell $n+1$ and those at unit cell
$n$ 
\begin{eqnarray}
\mathbf{L}(n+1) & = & \mathbb{T}(\epsilon,k_{x}).\mathbf{L}(n),\label{eq:BulkPassageMrel}
\end{eqnarray}
 where $\mathbf{L}(n)=[A(k_{x},n),B(k_{x},n)]^{T}$. The transfer matrix
$\mathbb{T}(\epsilon,k_{x})$ has the explicit form 
\begin{eqnarray}
\mathbb{T}(\epsilon,k_{x}) & = & -\frac{e^{-i\frac{k_{x}a}{2}}}{2\cos\Big(\frac{k_{x}a}{2}\Big)}\left[\begin{array}{cc}
1 & \frac{\epsilon}{t}\\
-\frac{\epsilon}{t} & 4\cos^{2}\Big(\frac{k_{x}a}{2}\Big)-\frac{\epsilon^{2}}{t^{2}}
\end{array}\right]\,.\label{eq:BulkPassageM}
\end{eqnarray}

As said in Section \ref{sec:GenDescrp}, the eigenvalues and eigenvectors
in the bulk are obtained from the diagonalization of the matrix $\mathbb{T}(\epsilon,k_{x})$.
In particular, eigenvalues of the transfer matrix which have unit
modulus, $\vert\lambda\vert^{2}=1$, correspond to Bloch solutions
propagating along the one-dimensional chain (a band state); eigenvalues
with a modulus that is different from one, $\vert\lambda\vert^{2}\neq1$,
correspond to evanescent Bloch solutions. These evanescent solutions
decrease with $n\to+\infty$ ($n\to-\infty$) when $\vert\lambda\vert^{2}<1$
($\vert\lambda\vert^{2}>1$).


\subsection{The defect}

\label{sec:JS-defect}

We now go on to write the TB equations at the defect (see Fig. \ref{fig_Scheme_1Dchain_doubleJSGB}),
and work them out in such a way that we can write a boundary condition
in the form given by Eq. (\ref{eq:DfctPassageRelGen}).

The TB equations at the defect (see Fig. \ref{fig_Scheme_1Dchain_doubleJSGB})are:
\begin{subequations} \label{eq:DefectTB} 
\begin{eqnarray}
\epsilon A(k_{x},1) & = & -(t')^{*}B(k_{x},0)-tB(k_{x},1);\label{eq:DefectTB1}\\
\epsilon B(k_{x},0) & = & -tD(k_{x})-t'A(k_{x},1);\\
\epsilon D(k_{x}) & = & -t\big(A(k_{x},0)+B(k_{x},0)\big)-2\xi t\cos(k_{x}a)D(k_{x});\\
\epsilon A(k_{x},0) & = & -(t')^{*}B(k_{x},-1)-tD(k_{x});\\
\epsilon B(k_{x},-1) & = & -t'A(k_{x},0)-tA(k_{x},-1)\,.\label{eq:DefectTB5}
\end{eqnarray}
 \end{subequations}

As we have mentioned above, the aim is to obtain an equation relating
the wave function at the two sides of the defect. With that in mind,
we solve each equation in (\ref{eq:DefectTB}) for the rightmost amplitude
appearing in it:\begin{subequations} \label{eq:DefectTB2} 
\begin{eqnarray}
B(k_{x},1) & = & -\frac{\epsilon}{t}A(k_{x},1)-\frac{(t')^{*}}{t}B(k_{x},0),\\
A(k_{x},1) & = & -\frac{\epsilon}{t'}B(k_{x},0)-\frac{t}{t'}D(k_{x}),\\
B(k_{x},0) & = & -\frac{\epsilon+2\xi t\cos(k_{x}a)}{t}D(k_{x})-A(k_{x},0),\\
D(k_{x}) & = & -\frac{\epsilon}{t}A(k_{x},0)-\frac{(t')^{*}}{t}B(k_{x},-1),\\
A(k_{x},0) & = & -\frac{\epsilon}{t'}B(k_{x},-1)-\frac{t}{t'}A(k_{x},-1).
\end{eqnarray}
 \end{subequations} It is convenient to write the above set of equations
in matrix form as \begin{subequations} \label{eq:DefectTB3} 
\begin{eqnarray}
\left[\begin{array}{c}
B(k_{x},1)\\
A(k_{x},1)
\end{array}\right] & = & -\left[\begin{array}{cc}
\frac{\epsilon}{t} & \frac{(t')^{*}}{t}\\
-1 & 0
\end{array}\right]\left[\begin{array}{c}
A(k_{x},1)\\
B(k_{x},0)
\end{array}\right],\\
\left[\begin{array}{c}
A(k_{x},1)\\
B(k_{x},0)
\end{array}\right] & = & -\left[\begin{array}{cc}
\frac{\epsilon}{t'} & \frac{t}{t'}\\
-1 & 0
\end{array}\right]\left[\begin{array}{c}
B(k_{x},0)\\
D(k_{x})
\end{array}\right],\\
\left[\begin{array}{c}
B(k_{x},0)\\
D(k_{x})
\end{array}\right] & = & -\left[\begin{array}{cc}
\frac{\epsilon}{t}+2\xi\cos(k_{x}a) & 1\\
-1 & 0
\end{array}\right]\left[\begin{array}{c}
D(k_{x})\\
A(k_{x},0)
\end{array}\right],\\
\left[\begin{array}{c}
D(k_{x})\\
A(k_{x},0)
\end{array}\right] & = & -\left[\begin{array}{cc}
\frac{\epsilon}{t} & \frac{(t')^{*}}{t}\\
-1 & 0
\end{array}\right]\left[\begin{array}{c}
A(k_{x},0)\\
B(k_{x},-1)
\end{array}\right],\\
\left[\begin{array}{c}
A(k_{x},0)\\
B(k_{x},-1)
\end{array}\right] & = & -\left[\begin{array}{cc}
\frac{\epsilon}{t'} & \frac{t}{t'}\\
-1 & 0
\end{array}\right]\left[\begin{array}{c}
B(k_{x},-1)\\
A(k_{x},-1)
\end{array}\right].
\end{eqnarray}
 \end{subequations}

It is now straightforward to write the boundary condition connecting
the two sides of the defect as 
\begin{eqnarray}
\mathbf{L}(1) & = & \mathbb{M}_{55}.\mathbf{L}(-1),\label{eq:JSTB_BC}
\end{eqnarray}
 where the matrix $\mathbb{M}_{55}$ is a $2\times2$ matrix defined
by 
\begin{eqnarray}
\mathbb{M}_{55} & = & R.\mathbb{N}_{1}(\epsilon,k_{x}).\mathbb{N}_{2}(\epsilon,k_{x}).\mathbb{N}_{3}(\epsilon,k_{x}).\mathbb{N}_{1}(\epsilon,k_{x}).\mathbb{N}_{2}(\epsilon,k_{x}).R^{T};\label{eq:JSmatrixM}
\end{eqnarray}
the matrix $R$ is the $\sigma_{x}$ Pauli matrix, used to interchange
rows $A$ and $B$. The matrices $\mathbb{N}_{1}$, $\mathbb{N}_{2}$, and $\mathbb{N}_{3}$,
after substituting $t'=t(1+e^{ik_{x}a})$, read \begin{subequations}
\label{eq:JSmatricesM} 
\begin{eqnarray}
\mathbb{N}_{1}(\epsilon,k_{x}) & = & -\left[\begin{array}{cc}
\frac{\epsilon}{t} & (1+e^{-ik_{x}a})\\
-1 & 0
\end{array}\right],\\
\mathbb{N}_{2}(\epsilon,k_{x}) & = & -\frac{1}{1+e^{ik_{x}a}}\left[\begin{array}{cc}
\frac{\epsilon}{t} & 1\\
-(1+e^{ik_{x}a}) & 0
\end{array}\right],\\
\mathbb{N}_{3}(\epsilon,k_{x}) & = & -\left[\begin{array}{cc}
\frac{\epsilon+2t\xi\cos(k_{x}a)}{t} & 1\\
-1 & 0
\end{array}\right],
\end{eqnarray}
 \end{subequations}

From Eqs. (\ref{eq:JSTB_BC})-(\ref{eq:JSmatricesM}) we conclude
that matrix $\mathbb{M}_{55}$, and consequently the boundary condition
imposed by the defect, depend both on the energy, $\epsilon$, and
on the longitudinal momentum, $k_{x}$.

\subsection{Computing the scattering coefficients}

\label{sec:JS-SctCoeffs}

Using the spectrum of modes allowed in the bulk of the one-dimensional
chain {[}obtained from the transfer matrix $\mathbb{T}(\epsilon,k_{x})${]}
and the boundary condition calculated above {[}see Eqs. (\ref{eq:JSTB_BC})-(\ref{eq:JSmatricesM}){]},
we can now solve completely the scattering problem from a \textit{pentagon-only}
defect line.

We consider the scattering process in which we have an incoming mode
from $n=-\infty$. In this situation, due to the presence of the defect
at $n=0$, there will be a reflected mode as well as a transmitted
one. For now, let us suppose to be working at positive energy and
around the $\nu=+1$ Dirac point {[}$\mathbf{K}_{+}=(4\pi/3a,0)${]}.

We choose to denote $\vert \Psi_{>}\rangle$ ($\vert\Psi_{<}\rangle$)
as the positively (negatively) moving mode of the transfer matrix
when we are around the $\mathbf{K}_{+}$ Dirac point. Furthermore,
the eigenvalue associated with this mode is going to be denoted by
$\lambda_{>}$ ($\lambda_{<}$). Therefore, we can write the wave function
on each side of the defect as \begin{subequations} \label{eq:WaveFunctions}
\begin{eqnarray}
\vert \Psi(n<0)\rangle & = & \lambda_{>}^{n+1}\vert \Psi_{>}\rangle+\rho\lambda_{<}^{n+1}\vert \Psi_{<}\rangle,\\
\vert \Psi(n>0)\rangle & = & \tau\lambda_{>}^{n-1}\vert \Psi_{>}\rangle,
\end{eqnarray}
 \end{subequations} where $\rho$ and $\tau$ stand for the reflection
and transmission scattering amplitudes, respectively.

The direction of propagation of each of the modes obtained by diagonalization
of the transfer matrix, can be determined by calculating the corresponding
current. It is shown in \ref{app:ConsvCurr} that, for graphene, the
current operator along the equivalent 1D-chain is $t\sigma_{y}/\hbar$;
in Fig. \ref{fig:CurrModsPristine} we plot the current associated
with each one of the transfer matrix's modes. For different choices
of $\epsilon$ and $k_{x}$ the incoming, reflected and transmitted
modes must be chosen accordingly with the direction of propagation of
the current.

In order to determine the scattering coefficients $\rho$ and $\tau$,
following the procedure detailed in Section \ref{sec:GenDescrp},
we write Eq.~(\ref{eq:BCGen_PropBasis}) for this specific case,
assuming an ordering of the modes as $\{\vert \Psi_{>}\rangle,\vert \Psi_{<}\rangle\}$;
the resulting equation is 
\begin{eqnarray}
\left[\begin{array}{c}
\tau\\
0
\end{array}\right]_{n=+1} & = & U^{-1}\mathbb{M}_{55}U\left[\begin{array}{c}
1\\
\rho
\end{array}\right]_{n=-1}.\label{eq:JSTB_BC2}
\end{eqnarray}
 Solving the above linear system is trivial, resulting in 
\begin{subequations} \label{eq:TBcoeffs} 
\begin{eqnarray}
\rho & = & -\frac{\big(U^{-1}\mathbb{M}_{55}U\big)_{21}}{\big(U^{-1}\mathbb{M}_{55}U\big)_{22}},\\
\tau & = & \frac{\det\left(U^{-1}\mathbb{M}_{55}U\right)}{\big(U^{-1}\mathbb{M}_{55}U\big)_{22}}. \label{eq:TBcoeffs-b}
\end{eqnarray}
 \end{subequations}

\subsection{The transmittance}

\label{sec:JS-transmtt}

\begin{figure}[htp!]
  \centering
  \includegraphics[clip,width=0.95\textwidth]{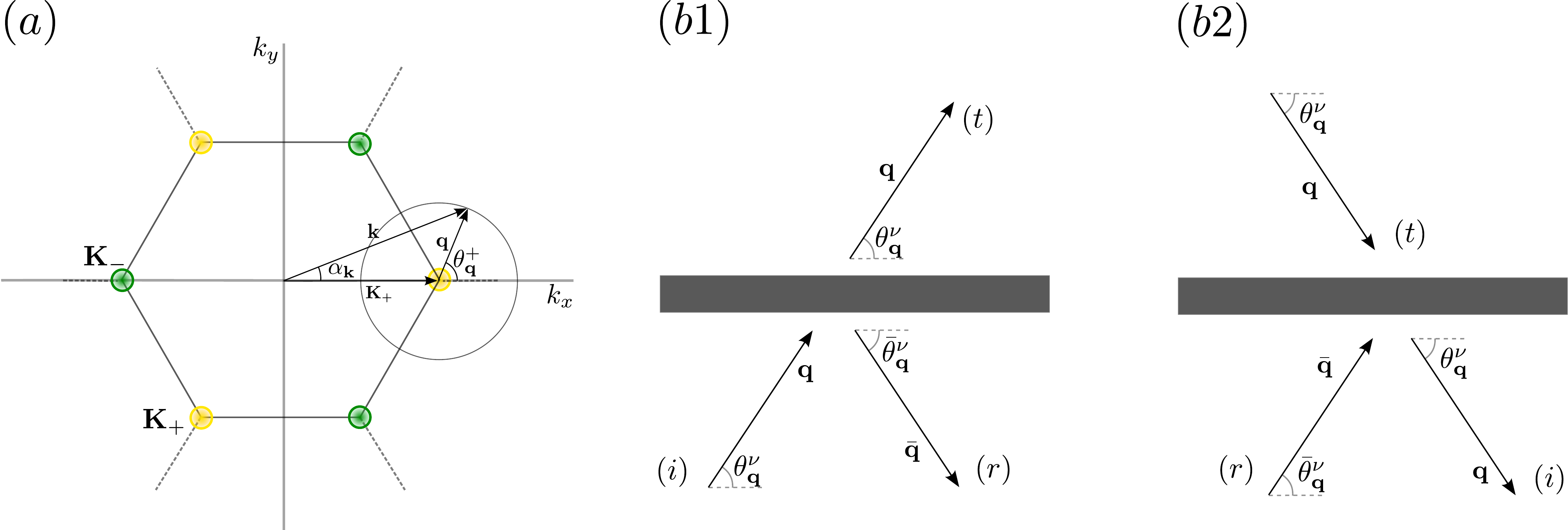}
  \caption{(Color online) (a) Pristine graphene FBZ where the incident
    wave vector, $\mathbf{k}$, is signaled, as well as the vector
    $\mathbf{q}=\mathbf{k}-\mathbf{K}_{\nu}$.  (b) Schemes of the
    electron scattering across a defect line in $2$D for low energy
    (wave vector around the Dirac cone) in two different cases: (b1)
    $\epsilon>0$; (b2) $\epsilon<0$. The labels $(i)$, $(r)$ and $(t)$
    stand respectively for the incident, reflected and transmitted
    modes' wave-vectors.}
  \label{fig:FBZ-Incd}
\end{figure}

In the context of graphene, we are mostly interested in the low-energy
region of the spectrum, that is close to the Dirac points. Therefore,
in what concerns the electronic transport, it is a natural choice
to plot the transmittance, $T=\vert\tau\vert^{2}$ in terms of $\mathbf{q}=\mathbf{k}-\mathbf{K}_{\nu}$,
where $\mathbf{k}$ is measured from the center of the zone {[}see
Fig. \ref{fig:FBZ-Incd}{]}.

In the low-energy limit, the graphene quasi-particles impinging on
the defect line are massless Dirac fermions with a wave-vector $\mathbf{q}$,
which defines their propagation direction. Therefore, in this limit,
it is intuitive to refer the transmittance to the angle $\theta$
that $\mathbf{q}$ makes with the defect line. As a consequence, and
despite the fact that we are not necessarily working at low energies,
we will choose to express the transmittance (obtained from the TB
model) in terms of $\epsilon$ and $\theta$, instead of doing so
in terms of $\epsilon$ and $k_{x}$. In \ref{app:IncdAngle} we show
how starting from the expressions of $T(\epsilon,k_{x})$ we can obtain
the transmittance in terms of the energy and the angle $\theta$,
namely $T(\epsilon,\theta)$. 

In Fig. \ref{fig:CurrModsPristine} it can be seen that the two modes
of the transfer matrix have opposite directions of propagation. The
figure also shows that when the sign of the energy is changed, the
direction of propagation of the modes is reversed. Suppose that for
a given pair $(\epsilon,k_{x})$, the mode identified by $\theta$
(or $+q_{y}$) is the one with positive direction of propagation.
Therefore, if we make the change $\epsilon\to-\epsilon$, then the
mode propagating in the positive direction along the one-dimensional
chain is now the one with $-\theta$ (or $-q_{y}$) {[}compare panels
(b1) and (b2) of Fig. \ref{fig:FBZ-Incd}{]}.

As a consequence, and for the sake of comparison between scattering
processes occurring at positive energy and those happening at negative
energy, we will plot the transmittance against the angle $\theta'=\theta$
when $\epsilon>0$, and against the angle $\theta'=-\theta$ when
$\epsilon<0$.

\begin{figure}[htp!]
  \centering
  \includegraphics[clip,width=0.65\textwidth]{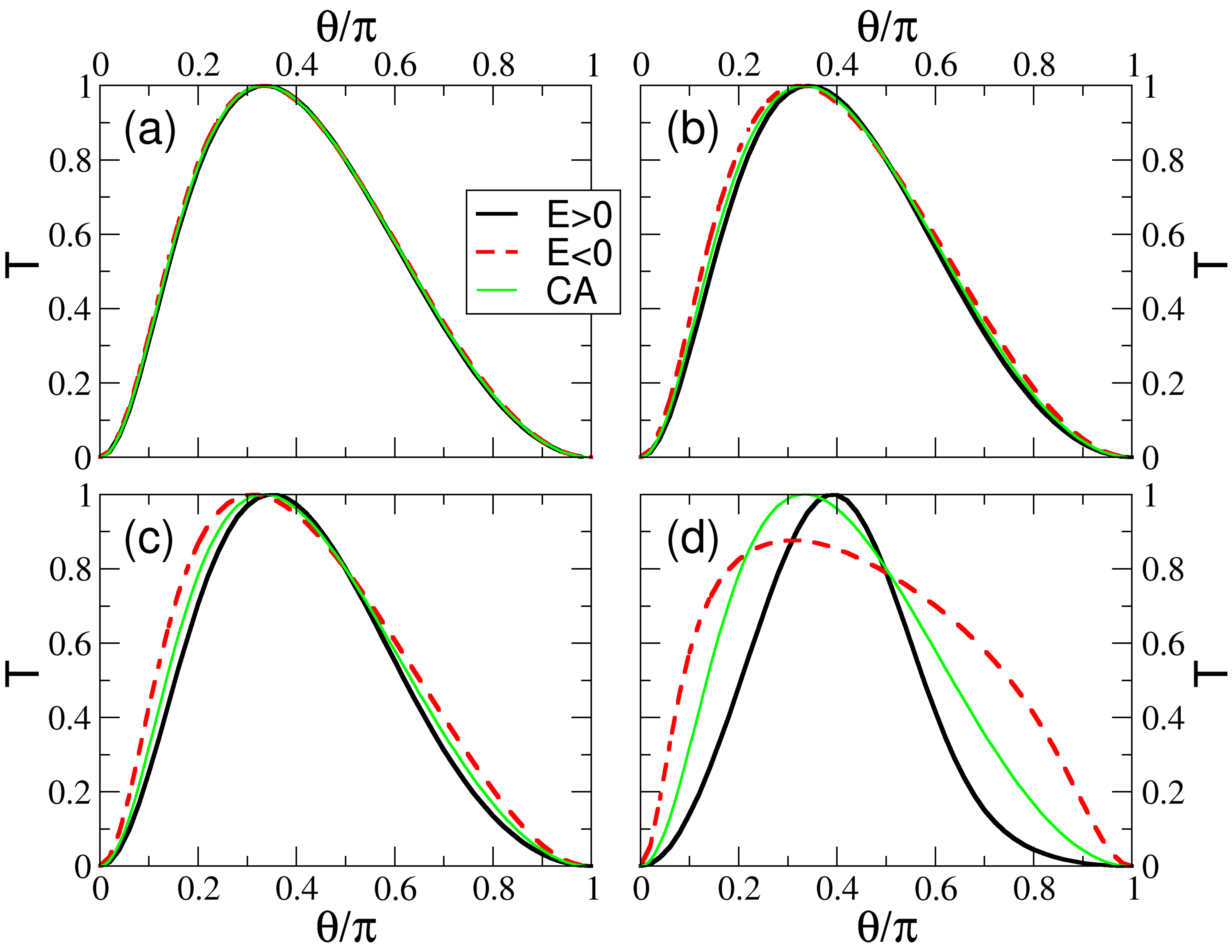}
  \caption{(Color online) Transmittance across the
    \textit{pentagon-only} defect line as function of the incoming
    angle $\theta$ between $\mathbf{q}$ (of the incoming particle) and
    the barrier. Remember that for negative energies, the
    transmittance is plotted against the angle $-\theta$.  The hopping
    parameter at the defect, $\xi$, was fixed to $\xi=1$ in this set
    of plots. The transmittance is plotted for several different
    modulus of the energy. These are: (a) $\vert\epsilon/t\vert=0.01$;
    (b) $\vert\epsilon/t\vert=0.05$; (c) $\vert\epsilon/t\vert=0.1$;
    (d) $\vert\epsilon/t\vert=0.5$. Positive energies are represented
    by the black full lines, while the negative ones are represented
    by the dashed red lines. The green full lines stand for the
    continuum low-energy result (that is energy-independent) as
    obtained in Ref. \cite{Rodrigues_PRB:2012}.  Only the curves
    associated with the Dirac point $\mathbf{K}_{+}$ are
    represented. Those for the Dirac point $\mathbf{K}_{-}$ are
    obtained from the former by a reflection of these over the axis
    $\theta=\pi/2$.}
  \label{fig_func_theta_pos_neg_E}
\end{figure}

In Fig. \ref{fig_func_theta_pos_neg_E} we represent the transmittance
$T=\vert\tau\vert^{2}$ as function of the angle $\theta$ (when $\epsilon>0$
and $-\theta$ when $\epsilon<0$), made between $\mathbf{q}$ of
the incoming particle and the barrier {[}see Fig. \ref{fig:FBZ-Incd}{]}.
In the different panels of Fig. \ref{fig_func_theta_pos_neg_E}, we
present the transmittance curves for several values of the energy.
The curves in the latter figure refer to the Dirac point $\mathbf{K}_{+}$.
Those referring to the Dirac point $\mathbf{K}_{-}$, are mirror symmetric
to the former ones relatively to the axis $\theta=\pi/2$. 

As one can see in Fig. \ref{fig_func_theta_pos_neg_E}(a) and (b), the TB low-energy result 
(black full line and red dashed line) is in good accordance with that obtained in Ref. 
\cite{Rodrigues_PRB:2012} using the continuum approximation (green full line).

The transmittance profile arising from the {\it pentagon-only} defect line is essentially controlled by
the values of the hopping parameter $\xi$ at the defect line [see Eqs. (\ref{eq:TBcoeffs})]. Modifications 
of this hopping parameter, result in very different transmittance behaviors: the transmission amplitude for
a given angle, $\theta$, and a given energy, $\epsilon$, is typically very different for distinct values of 
$\xi$; in particular, the existence and position of an angle with perfect transmittance is strongly dependent
on $\xi$.

Looking at Eq. (\ref{eq:TBcoeffs-b}) one can easily concludes that it is its denominator that controls the 
transmittance, since, from Eq. (\ref{eq:CurrConsvBCcondt}) and Eq. (\ref{eq:zz_ConsCurr}), we have that 
$\vert \det (U^{-1} \mathbb{M}_{5 5} U) \vert^{2} = 1$. Nevertheless, the analysis of the mathematical 
expression of $(U^{-1} \mathbb{M}_{5 5} U)_{2 2}$ gives us little physical intuition over the origin of an 
angle with perfect transmittance for a range of values of the hopping parameter $\xi$. Therefore, we will make use 
of the continuum low-energy description of this system, developed in Ref. \cite{Rodrigues_PRB:2012}, to 
investigate this feature of the transmittance.

As argued in Ref. \cite{Rodrigues_PRB:2012}, in the continuum low-energy limit, one can see the defect line as 
a strip of width $W$, where there is a general local potential, $\hat{V}(y)=V_{s} \mathbb{I}+V_{x} \sigma_{x}
+V_{y}\sigma_{y}
+V_{z}\sigma_{z}$. In the case of a {\it pentagon-only} defect line, one can show (see \ref{app:CAtransm}) that 
the general local potential is of the form: $V_{s} \neq 0 \neq V_{x}$ and $V_{y} = 0 = V_{z}$.
Inside the strip, the scalar potential term, $V_{s}$, changes the direction of propagation of the massless Dirac 
fermion, but keeps its spin aligned with its momentum (here and in the following paragraphs "spin" refers to the 
sub-lattice pseudo-spin degree of freedom). In contrast, the term $V_{x}$ has the effect 
of not only changing the direction of propagation of the fermion inside the strip, but also of misaligning its 
spin and its momentum (see \ref{app:CAtransm}). When $\epsilon, q_{x} \to 0$, the fermions inside the strip 
all propagate in the same direction, with the same misalignment between their spin and 
their momentum, and, therefore, share a common spin direction, at an angle $\alpha = \arccos (\xi/2)$ with the
line defect.

At very low energies, fermions incident on the strip at an angle $\alpha$
have their spin already aligned with the spin direction inside the strip, and thus, their wave-functions 
outside and inside the strip can be matched without a reflected wave; the strip will be {\it invisible} to
them and they will be totally transmitted across the defect line. Fermions at a different angle of incidence 
will have their spin outside and inside the strip misaligned, and thus will only be partly transmitted by 
the defect line. As a consequence, in the case of a {\it pentagon-only} defect line, the angle of perfect 
transmission (at very low energies) is exactly given by $\alpha = \arccos (\xi/2)$ [see Fig. 
\ref{fig_func_theta_pos_neg_E}(a)]. We see that this mechanism is entirely identical to that of  perfect 
normal transmission across a barrier with a scalar potential \cite{Katsnelson_NatPhys:2006}; the difference 
lies only with the presence of a $V_{x}$ term, which induces misalignment between spin and momentum, and 
changes the direction along which spin alignment inside and outside the barrier occurs.

In the cases where $\vert \xi \vert > 2$, the wave-number associated with the fermions propagating inside the 
strip, $\widetilde{q}_{y}$, becomes imaginary: its wave-function is evanescent inside the strip. Moreover, 
$\alpha$ becomes imaginary and thus, the spin of the fermion has a non-zero $z$ component. The fermions
outside and inside the strip can never have their spins perfectly aligned. Their wave-functions cannot
match without a reflected component, and there is no incident direction with perfect transmittance.

\section{Scattering from a $zz(5757)$ and a $zz(558)$ defect line}

\label{sec:zz5757-zz558}

\begin{figure}[!htp]
  \centering
  \includegraphics[width=0.49\textwidth]{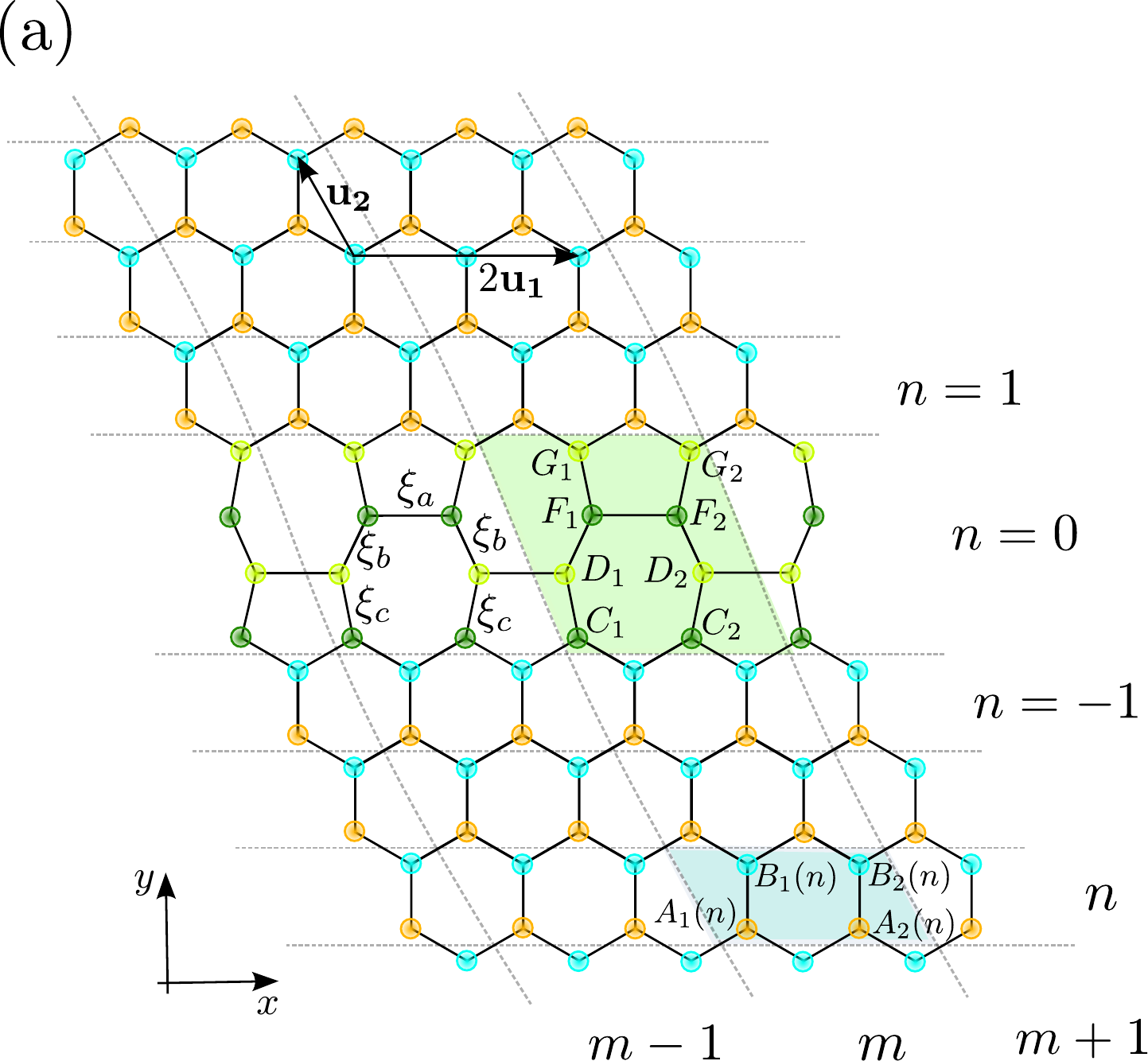}
  \includegraphics[width=0.49\textwidth]{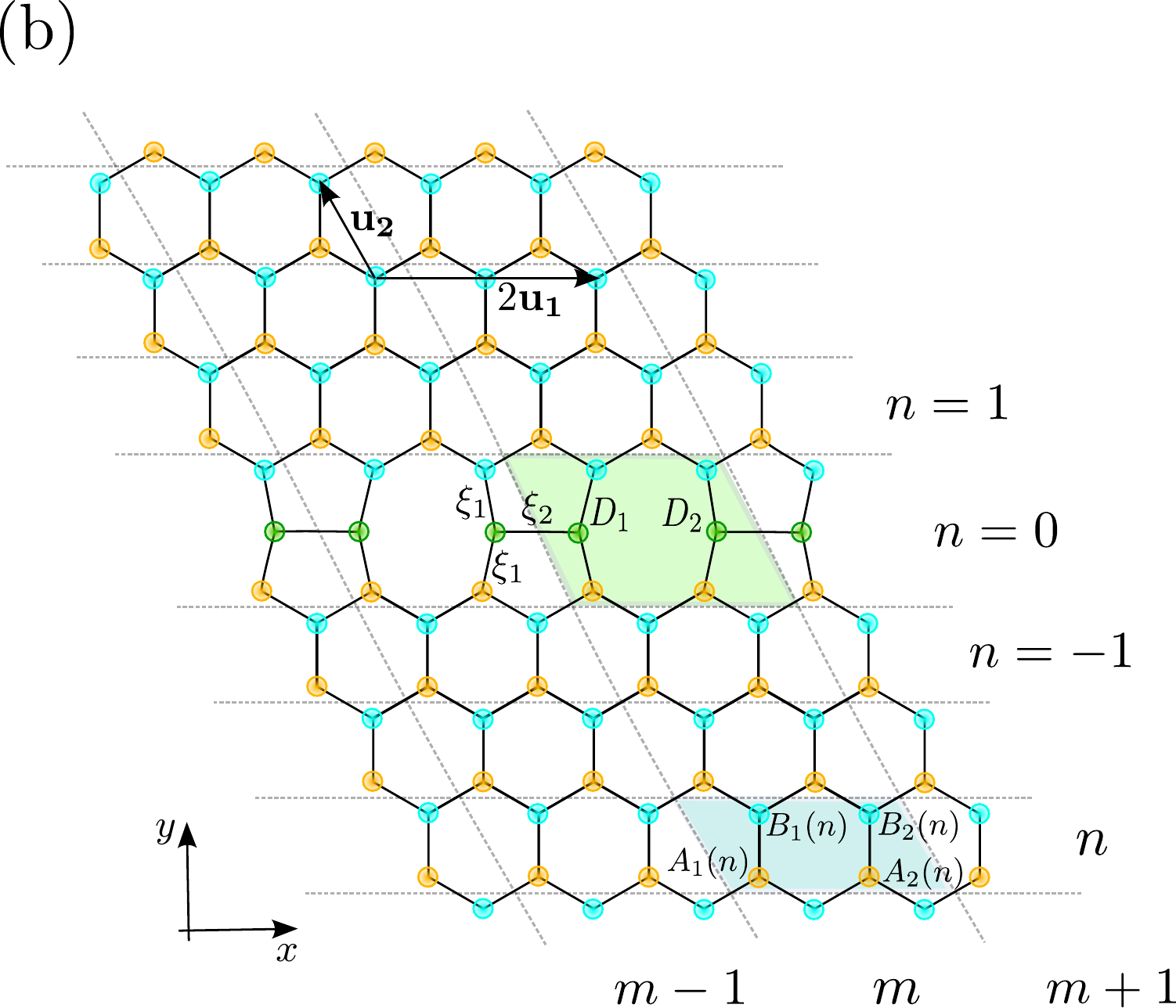}
  \caption{(Color online) Panel (a): Scheme of a $zz(5757)$ defect
    line. Panel (b): Scheme of a $zz(558)$ defect line.}
  \label{fig:Scheme_zz5757-zz558}
\end{figure}

In this section we consider two other types of extended defect lines,
namely the $zz(5757)$ \cite{Rodrigues_PRB:2012} and the $zz(558)$
defect lines \cite{Lahiri_NatureNanotech:2010,PRLDefect,PhysLettADefect}
{[}see Fig. \ref{fig:Scheme_zz5757-zz558}{]}. These defect lines
are more realistic, albeit somewhat more complex to treat. We will
see that both these defect lines give rise to a duplication of the
unit cell. This originates a folding of the Brillouin zone, that brings
to play two additional scattering modes. Despite the fact that at
low-energies the latter happen to be evanescent modes, they must be
considered in the computation of the scattering amplitudes. 

The treatments of both the $zz(5757)$ and the $zz(558)$ defect lines
are very similar to each other. The only difference between them is
on the microscopic details of each defect line. As a consequence,
and for the sake of definiteness, we shall first study electron scattering
from a $zz(5757)$ defect line. We will employ again the procedure
described in Section \ref{sec:GenDescrp}, which was previously applied
to the study of the \textit{pentagon-only} defect line (see Section
\ref{sec:doublepentagonGB}). Afterwards, we will also compute the
boundary condition arising from the $zz(558)$ defect line and we
present the results for both the $zz(5757)$ and the $zz(558)$ cases.

\subsection{Formulating the problem}

\label{sec:zzD-Form}

Let us thus concentrate on the study of a graphene layer with a $zz(5757)$
defect line along the zigzag direction. As can be seen in Fig. \ref{fig:Scheme_zz5757-zz558}(a),
the $zz(5757)$ defect line has a periodicity twice as large as that
of the \textit{pentagon-only} defect line {[}see Fig. \ref{fig_Scheme_doubleJS_GB}{]}.
Therefore, its unit cell will necessarily be two times bigger in the
defect direction, when compared with the unit cell of pristine graphene.

\begin{figure}[!htp]
 \centering \includegraphics[width=0.75\textwidth]{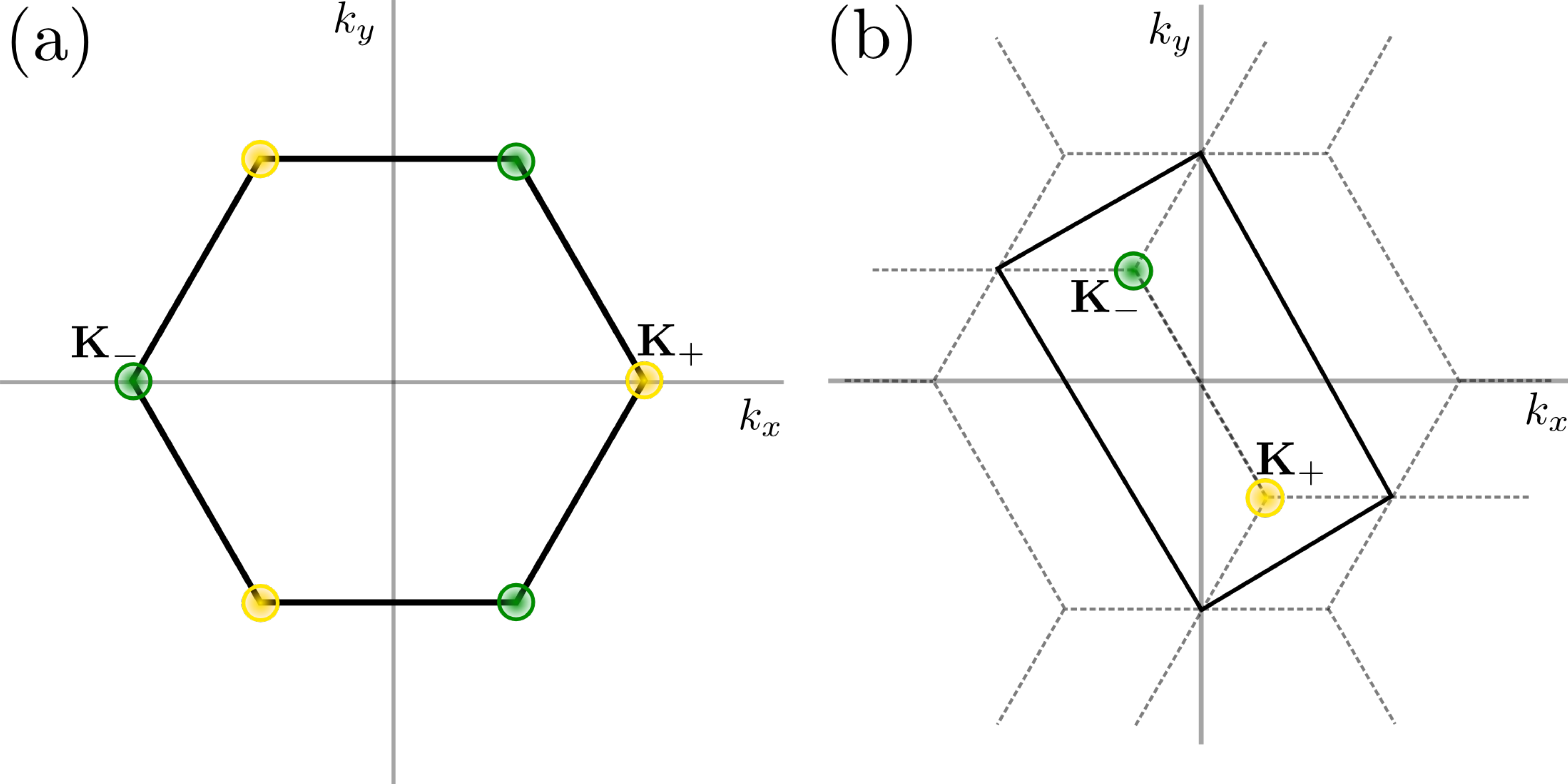}
\caption{(Color online) Panel (a): Scheme of the FBZ of pristine graphene.
Panel (b): Scheme of the FBZ of graphene with a doubled unit cell
in $\mathbf{u}_{1}$ direction.}

\label{fig:FBZs} 
\end{figure}

Then, the most natural choice for the lattice vectors is $2\mathbf{u}_{1}$
and $\mathbf{u}_{2}$ {[}where $\mathbf{u}_{1}$ and $\mathbf{u}_{2}$
were defined in Eq. (\ref{eq:LatticeVecs}){]}. As a consequence the
unit cell of bulk graphene will be twice as big in the $\mathbf{u}_{1}$
direction. Therefore, the First Brillouin Zone (FBZ) will be folded
in comparison with the one of pristine graphene (see Fig. \ref{fig:FBZs}).
The Dirac points will then be located at $\mathbf{K}_{\nu}=\nu\pi/(3a)(1,-\sqrt{3})$
(see \ref{app:IncdAngle}), where again, $\nu=\pm1$ identifies the
Dirac point. 

Let us now write the TB Hamiltonian of the graphene layer with a $zz(5757)$
defect line {[}see Fig. \ref{fig:Scheme_zz5757-zz558}(a){]}. As before,
we can separate the Hamiltonian in three distinct parts: $\hat{H}=\hat{H}^{U}+\hat{H}^{D}+\hat{H}^{L}$,
where $\hat{H}^{U}$ ($\hat{H}^{L}$) stands for the Hamiltonian above
(below) the defect line, while the remaining term, $\hat{H}^{D}$,
describes the defect line itself. In second quantization the explicit
forms of $\hat{H}^{U}$ and $\hat{H}^{L}$ read 
\begin{eqnarray}
H^{U(L)} & = & -t\sum_{m}\sum_{n}\Bigg[\bigg(b_{1}^{\dagger}(m,n)+b_{1}^{\dagger}(m,n-1)+b_{2}^{\dagger}(m-1,n-1)\bigg)a_{1}(m,n)\nonumber \\
 &  & +\bigg(b_{2}^{\dagger}(m,n)+b_{1}^{\dagger}(m,n-1)+b_{2}^{\dagger}(m,n-1)\bigg)a_{2}(m,n)+h.c.\Bigg]\,,\label{eq:zzD-Hbulk}
\end{eqnarray}
where for $H^{U}$ ($H^{L}$) for $n\geq1$ ($n\leq-1$).

Similarly, the term describing the TB Hamiltonian at the defect line
reads 
\begin{eqnarray}
H^{D} & = & -t\sum_{m}\Bigg[\bigg(\xi_{c}d_{1}^{\dagger}(m)+b_{1}^{\dagger}(m,-1)+b_{2}^{\dagger}(m-1,-1)\bigg)c_{1}(m)\nonumber \\
 &  & +\bigg(\xi_{c}d_{2}^{\dagger}(m)+b_{1}^{\dagger}(m,-1)+b_{2}^{\dagger}(m,-1)\bigg)c_{2}(m)+\xi_{a}d_{1}^{\dagger}(m)d_{2}(m-1)\nonumber \\
 &  & +\xi_{a}f_{1}^{\dagger}(m)f_{2}(m)+\xi_{b}d_{1}^{\dagger}(m)f_{1}(m)+\xi_{b}d_{2}^{\dagger}(m)f_{2}(m)+\bigg(\xi_{c}f_{1}^{\dagger}(m)\nonumber \\
 &  & +a_{1}^{\dagger}(m,1)+a_{2}^{\dagger}(m,1)\bigg)g_{1}(m)+\bigg(\xi_{c}f_{2}^{\dagger}(m)+a_{1}^{\dagger}(m+1,1)\nonumber \\
 &  & +a_{2}^{\dagger}(m,1)\bigg)g_{2}(m)+h.c.\Bigg]\,,\label{eq:zzD-HDfct}
\end{eqnarray}
where $\xi_{a}$, $\xi_{b}$ and $\xi_{c}$ stand for the renormalizations
of the hopping amplitudes at the defect {[}see Fig. \ref{fig:Scheme_zz5757-zz558}(a){]}.

\begin{figure}[htp!]
 \centering \includegraphics[clip,width=0.95\textwidth]{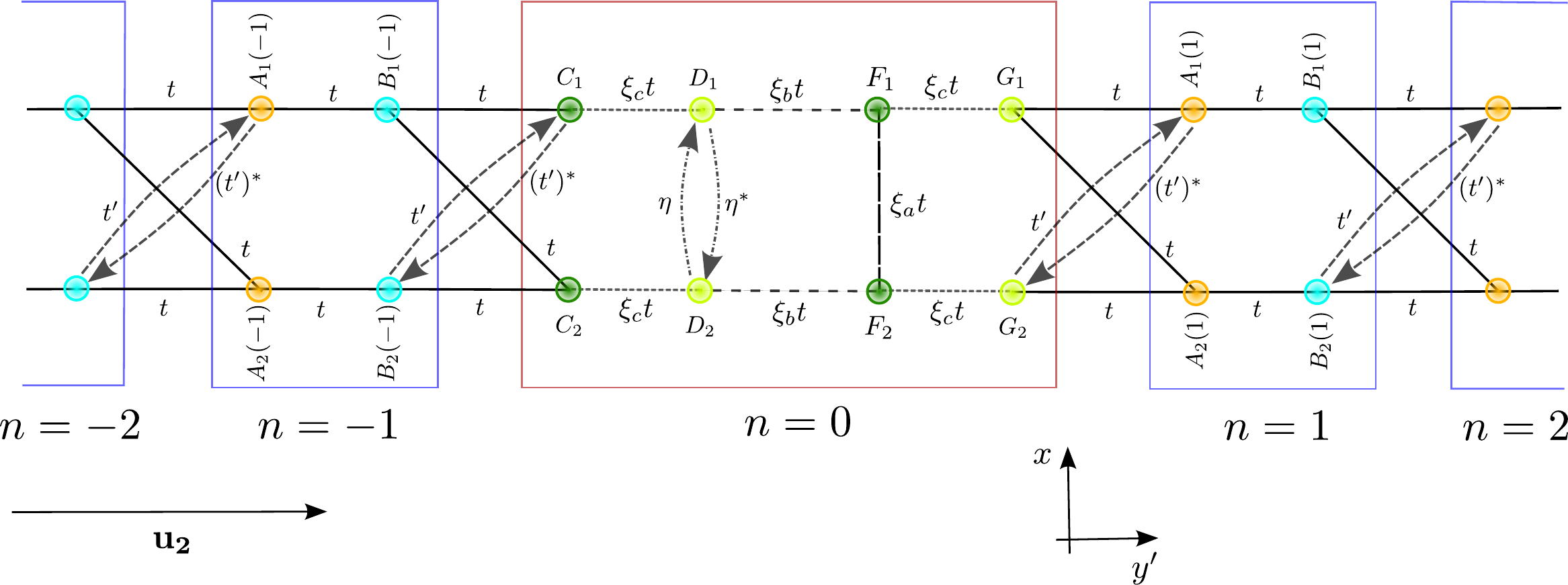}
\caption{(Color online) Effective one-dimensional chain obtained after Fourier
transforming the Hamiltonian of graphene with a $zz(5757)$ defect
line. The hopping parameters $t'$ and $\eta$, are short hands for
$t'=te^{-i2k_{x}a}$ and for $\eta=t\xi_{a}e^{-i2k_{x}a}$. The parameters
$\xi_{a}$, $\xi_{b}$ and $\xi_{c}$ stand for the hopping renormalizations
appearing in Fig. \ref{fig:Scheme_zz5757-zz558}(a).}

\label{fig:zz5757-1D} 
\end{figure}

Again the system is invariant under translations $\mathbf{r}=2m\mathbf{u}_{1}$
(where $m$ is an integer). Consequently, we can make use of Bloch
theorem and Fourier transform the Hamiltonian along this direction,
diagonalizing it with respect to the variable $m$. As expected, the
resulting Hamiltonian turns out to be equivalent to the Hamiltonian
describing a quasi-one-dimensional chain, with four atoms per unit
cell and a defect at its center (see Fig. \ref{fig:zz5757-1D}). 

Therefore, we can write the Hamiltonian of the effective one-dimensional
chain as $H(k_{x})=H^{U}(k_{x})+H^{D}(k_{x})+H^{L}(k_{x})$. The part
above and below the defect read 
\begin{eqnarray}
H^{U(L)}(k_{x}) & = & -t\sum_{n}\Bigg[\bigg(b_{1}^{\dagger}(k_{x},n)+b_{1}^{\dagger}(k_{x},n-1)+e^{i2k_{x}a}b_{2}^{\dagger}(k_{x},n-1)\bigg)\nonumber \\
 &  & \times a_{1}(k_{x},n)+\bigg(b_{2}^{\dagger}(k_{x},n)+b_{1}^{\dagger}(k_{x},n-1)+b_{2}^{\dagger}(k_{x},n-1)\bigg)\nonumber \\
 &  & \times a_{2}(k_{x},n)+h.c.\Bigg]\,,\label{eq:zzD-Hbulk-kx}
\end{eqnarray}
while the part corresponding to the defect reads 
\begin{eqnarray}
H^{D}(k_{x}) & = & -t\Bigg[\bigg(\xi_{c}d_{1}^{\dagger}(k_{x})+b_{1}^{\dagger}(k_{x},-1)+e^{i2k_{x}a}b_{2}^{\dagger}(k_{x},-1)\bigg)c_{1}(k_{x})+\bigg(\xi_{c}d_{2}^{\dagger}(k_{x})\nonumber \\
 &  & +b_{1}^{\dagger}(k_{x},-1)+b_{2}^{\dagger}(k_{x},-1)\bigg)c_{2}(k_{x})+\xi_{a}e^{-i2k_{x}a}d_{1}^{\dagger}(k_{x})d_{2}(k_{x})\nonumber \\
 &  & +\xi_{a}f_{1}^{\dagger}(k_{x})f_{2}(k_{x})+\xi_{b}d_{1}^{\dagger}(k_{x})f_{1}(k_{x})+\xi_{b}d_{2}^{\dagger}(k_{x})f_{2}(k_{x})+\bigg(\xi_{c}f_{1}^{\dagger}(k_{x})\nonumber \\
 &  & +a_{1}^{\dagger}(k_{x},1)+a_{2}^{\dagger}(k_{x},1)\bigg)g_{1}(k_{x})+\bigg(\xi_{c}f_{2}^{\dagger}(k_{x})+e^{-i2k_{x}a}a_{1}^{\dagger}(k_{x},1)\nonumber \\
 &  & +a_{2}^{\dagger}(k_{x},1)\bigg)g_{2}(k_{x})+h.c.\Bigg]\,.\label{eq:zzD-HDfct-kx}
\end{eqnarray}

As was stated in Section \ref{sec:JS-form}, the TB equations of this
system are obtained from its Schr\"{o}dinger equation, $\hat{H}(k_{x})\vert\mu,k_{x}\rangle=\epsilon_{\mu,k_{x}}\vert\mu,k_{x}\rangle$.
In this last equation, $\hat{H}(k_{x})$ stands for the Hamiltonian
of the effective one-dimensional chain given by Eqs. (\ref{eq:zzD-Hbulk-kx})-(\ref{eq:zzD-HDfct-kx}).
At the same time, the eigenstate $|\mu,k_{x}\rangle$ can be expressed
as a linear combination of the site amplitudes along the one-dimensional
chain: 
\begin{eqnarray}
\vert\mu,k_{x}\rangle & = & \sum_{i=1}^{2}\sum_{n\neq0}\Big[A_{i}(k_{x},n)\vert a_{i};k_{x},n\rangle+B_{i}(k_{x},n)\vert b_{i};k_{x},n\rangle\Big]\nonumber \\
 &  & +\sum_{i=1}^{2}\Big[C_{i}(k_{x})\vert c_{i};k_{x}\rangle+D_{i}(k_{x})\vert d_{i};k_{x}\rangle+F_{i}(k_{x})\vert f_{i};k_{x}\rangle\nonumber \\
 &  & +G_{i}(k_{x})\vert g_{i};k_{x}\rangle\Big],\label{eq:zzD-SchrState}
\end{eqnarray}
 where the $\vert z_{i};k_{x},n\rangle=\hat{z}_{i}^{\dagger}(k_{x},n)\vert0\rangle$
stand for the one-particle states at the atom $Z_{i}$ of unit cell
$n$ of the one-dimensional chain {[}with $z=a,b$ on the bulk ($n\neq0$)
and $z=c,d,f,g$ at the defect ($n=0$){]}.


\subsection{Bulk properties}

\label{sec:zzD-bulk}

From the above discussion, we can readily write the TB equations in
the bulk ($n\neq0$) of the quasi-one-dimensional chain. They read
\begin{subequations} \label{eq:zzD-TBbulkEqs} 
\begin{eqnarray}
\epsilon A_{1}(k_{x},n) & = & -tB_{1}(k_{x},n)-tB_{1}(k_{x},n-1)-te^{-i2k_{x}a}B_{2}(k_{x},n-1)\,,\label{eq:zzD-TBbulkEqs1}\\
\epsilon B_{1}(k_{x},n) & = & -tA_{1}(k_{x},n)-tA_{1}(k_{x},n+1)-tA_{2}(k_{x},n+1)\,,\label{eq:zzD-TBbulkEqs2}\\
\epsilon A_{2}(k_{x},n) & = & -tB_{2}(k_{x},n)-tB_{1}(k_{x},n-1)-tB_{2}(k_{x},n-1)\,,\label{eq:zzD-TBbulkEqs3}\\
\epsilon B_{2}(k_{x},n) & = & -tA_{2}(k_{x},n)-te^{i2k_{x}a}A_{1}(k_{x},n+1)-tA_{2}(k_{x},n+1)\,.\label{eq:zzD-TBbulkEqs4}
\end{eqnarray}
\end{subequations}

As before, we can write a recurrence relation between the amplitudes
of the unit cell located at position $n$ and those of the unit cell
located at position $n+1$. Following a procedure similar to that
leading to Eqs. (\ref{eq:BulkPassageMrel}) and (\ref{eq:BulkPassageM}),
one obtains 
\begin{eqnarray}
\mathbf{L}(n+1) & = & \mathbb{T}(\epsilon,k_{x}).\mathbf{L}(n),\label{eq:zzD-BulkPassageMrel}
\end{eqnarray}
 where $\mathbf{L}(n)=[A_{1}(k_{x},n),B_{1}(k_{x},n),A_{2}(k_{x},n),B_{2}(k_{x},n)]^{T}$,
and the transfer matrix, $\mathbb{T}(\epsilon,k_{x})$, has the following
form 
\begin{eqnarray}
\mathbb{T}(\epsilon,k_{x}) & = & -\frac{1}{1-e^{i2k_{x}a}}\left[\begin{array}{cccc}
-1 & 1 & -\epsilon & \epsilon\\
e^{i2k_{x}a} & -1 & \frac{\epsilon}{e^{-i2k_{x}a}} & -\epsilon\\
\epsilon & -\epsilon & w(\epsilon,k_{x}) & -w(\epsilon,-k_{x})\\
\frac{-\epsilon}{e^{-i2k_{x}a}} & \epsilon & \frac{-w(\epsilon,-k_{x})}{e^{-i2k_{x}a}} & w(\epsilon,k_{x})
\end{array}\right],\label{eq:zzD-PassMat1}
\end{eqnarray}
 where $w(\epsilon,k_{x})=-1+e^{i2k_{x}a}+\epsilon^{2}$.

If we take into account the nature of the folding of the FBZ, we can
make a basis change that will turn the transfer matrix $\mathbb{T}(\epsilon,k_{x})$
into a block diagonal matrix (see \ref{app:BasisUncp}). Such a basis
change groups the modes in two distinct pairs, uncoupling the bulk
descriptions of each one of these pairs. We dub the modes of one of
those pairs as the $+$ modes, while the modes of the other are going
to be called $-$ modes. Each one of these pairs of modes is associated
with a pair of energy bands present in the folded FBZ. Note, for instance,
that when we position ourselves near the Dirac points, $\mathbf{K}_{\nu}=\nu\pi/3(1/a,-\sqrt{3}/a)$,
the $+$ modes turn out to be high in energy ($\epsilon_{+}\approx2t$),
while the $-$ modes are low in energy ($\epsilon_{-}\approx0$) \cite{Rodrigues_PRB:2011}.

The latter basis change reads 
\begin{eqnarray}
\overline{\mathbf{L}}(n) & = & \Lambda(k_{x})\cdot\mathbf{L}(n),\label{eq:UncpBasis}
\end{eqnarray}
where $\overline{\mathbf{L}}(n)=[A_{+}(k_{x},n),B_{+}(k_{x},n),A_{-}(k_{x},n),B_{-}(k_{x},n)]^{T}$,
while the matrix $\Lambda(k_{x})$, reads (see \ref{app:BasisUncp})
\begin{eqnarray}
\Lambda(k_{x}) & = & \frac{1}{\sqrt{2}}\left[\begin{array}{cccc}
1 & 0 & e^{-ik_{x}a} & 0\\
0 & 1 & 0 & e^{-ik_{x}a}\\
1 & 0 & -e^{-ik_{x}a} & 0\\
0 & 1 & 0 & -e^{-ik_{x}a}
\end{array}\right].\label{eq:zzD-BasisChangeMat}
\end{eqnarray}
 Applying this transformation to the transfer matrix (\ref{eq:zzD-PassMat1})
we obtain $\overline{\mathbb{T}}(\epsilon,k_{x})=\Lambda(k_{x})\mathbb{T}(\epsilon,k_{x})\Lambda^{-1}(k_{x})$,
where 
\begin{eqnarray}
\overline{\mathbb{T}}(\epsilon,k_{x}) & = & \left[\begin{array}{cc}
\mathbb{T}_{+}(\epsilon,k_{x}) & 0\\
0 & \mathbb{T}_{-}(\epsilon,k_{x})
\end{array}\right].\label{eq:zzD-doubledUCpassageM}
\end{eqnarray}
is block diagonal. In the above expression, $\mathbb{T}_{+}$ and
$\mathbb{T}_{-}$ are $2\times2$ transfer matrices associated with
the $+$ and the $-$ modes, respectively. These matrices are given
by: \begin{subequations} \label{eq:zzD-doubleUCpassageMats} 
\begin{eqnarray}
\mathbb{T}_{+}(\epsilon,k_{x}) & = & -\frac{e^{-i\frac{k_{x}a}{2}}}{2\cos\big(\frac{k_{x}a}{2}\big)}\left[\begin{array}{cc}
1 & \frac{\epsilon}{t}\\
-\frac{\epsilon}{t} & 4\cos^{2}\big(\frac{k_{x}a}{2}\big)-\frac{\epsilon^{2}}{t^{2}}
\end{array}\right];\\
\mathbb{T}_{-}(\epsilon,k_{x}a) & = & \frac{e^{-i\frac{k_{x}a}{2}}}{2i\sin\big(\frac{k_{x}a}{2}\big)}\left[\begin{array}{cc}
1 & \frac{\epsilon}{t}\\
-\frac{\epsilon}{t} & 4\sin^{2}\big(\frac{k_{x}a}{2}\big)-\frac{\epsilon^{2}}{t^{2}}
\end{array}\right].
\end{eqnarray}
 \end{subequations}

Again, as was said in Section \ref{sec:GenDescrp}, the eigenvalues
and eigenvectors of the transfer matrix $\overline{\mathbb{T}}$
give the Bloch solutions of the problem. Eigenvalues with unit modulus,
$\vert\lambda\vert^{2}=1$, correspond to Bloch solutions propagating
along the one-dimensional chain (a band state), while eigenvalues
with non-unit modulus, $\vert\lambda\vert^{2}\neq1$, correspond to
evanescent Bloch solutions {[}decreasing with $n\to+\infty$ ($n\to-\infty$)
when $\vert\lambda\vert^{2}<1$ ($\vert\lambda\vert^{2}>1$){]}.

Moreover, in the basis that uncouples $+$ and $-$ modes the matrices
$\mathbb{T}_{+}$ and $\mathbb{T}_{-}$ give the Bloch solutions associated
with each one of the $+$ and $-$ energy sectors. Again, these can
be either propagating or evanescent depending on the energy $\epsilon$
and momentum $k_{x}$. Of relevance is the case where the scattering
process occurs at a sufficiently small energy ($\epsilon\lesssim2t$)
so that we are sufficiently near the Dirac points. In this case, the
$+$ modes turn out to be evanescent while the $-$ ones are still
propagating. When we are at very low-energies, $\epsilon\approx0$,
the transfer matrices associated with each pair of modes read \begin{subequations}
\label{eq:zzD-doubleUCpassageMats-LowEn} 
\begin{eqnarray}
\mathbb{T}_{+}\Big(0,\nu\frac{\pi}{3a}\Big) & = & -e^{-i\nu\frac{\pi}{6}}\left[\begin{array}{cc}
\frac{1}{\sqrt{3}} & 0\\
0 & \sqrt{3}
\end{array}\right],\\
\mathbb{T}_{-}\Big(0,\nu\frac{\pi}{3a}\Big) & = & e^{-i\nu\frac{2\pi}{3}}\left[\begin{array}{cc}
1 & 0\\
0 & 1
\end{array}\right].
\end{eqnarray}
 \end{subequations} Thus, we can conclude that the $+$ modes are
evanescent either decreasing or increasing exponentially as $e^{-n\log\sqrt{3}}$
or as $e^{n\log\sqrt{3}}$ with increasing $n$. At the same time
the $-$ modes are propagating.

\subsection{The $zz(5757)$ defect}

\label{sec:zzD-zz5757}

It is important to note that the conclusions of Section \ref{sec:zzD-bulk}
apply not only to the case of the $zz(5757)$ defect line, but also
to the $zz(558)$ defect line. In fact, the reasoning pursued in Section
\ref{sec:zzD-bulk} describes the bulk of systems whose unit cell's
size in the direction of $\mathbf{u}_{1}$ is twice that of pristine
graphene. As we can readily conclude from Fig. \ref{fig:Scheme_zz5757-zz558},
both the $zz(5757)$ defect line and the $zz(558)$ one impose the
same duplication of the unit cell along the zigzag direction. As a
consequence, the bulk problem of these two systems is described by
the same transfer matrix relations, Eqs. (\ref{eq:zzD-PassMat1})-(\ref{eq:zzD-doubleUCpassageMats}).

As previously noted, what distinguishes electron scattering in the
$zz(5757)$ and in the $zz(558)$ defect lines are the microscopic
details of each one of these defects. Next we will show how to compute
the boundary condition relation for each one of these two defect lines.

Let us start by the case of the $zz(5757)$ defect. From Eqs. (\ref{eq:zzD-HDfct-kx})
and (\ref{eq:zzD-SchrState}) we write the TB equations at the $zz(5757)$
defect. They read \begin{subequations} \label{eq:zz5757-TBeqsDfct}
\begin{eqnarray}
-\frac{\epsilon}{t}\mathbf{A}(1) & = & W_{A}^{\dagger}\mathbf{G}(0)+\mathbf{B}(1),\\
-\frac{\epsilon}{t}\mathbf{G}(0) & = & \xi_{c}\mathbf{F}(0)+W_{A}\mathbf{A}(1),\\
-\frac{\epsilon}{t}\mathbf{F}(0) & = & \xi_{b}\mathbf{D}(0)+\xi_{c}\mathbf{G}(0)+\xi_{a}\sigma_{x}\mathbf{F}(0),\\
-\frac{\epsilon}{t}\mathbf{D}(0) & = & \xi_{c}\mathbf{C}(0)+\xi_{b}\mathbf{F}(0)+\xi_{a}\sigma_{x}'\mathbf{D}(0),\\
-\frac{\epsilon}{t}\mathbf{C}(0) & = & W_{A}^{\dagger}\mathbf{B}(-1)+\xi_{c}\mathbf{D}(0),\\
-\frac{\epsilon}{t}\mathbf{B}(-1) & = & \mathbf{A}(-1)+W_{A}\mathbf{C}(0),
\end{eqnarray}
 \end{subequations} where $\mathbf{Z}(n)=[Z_{1}(n),Z_{2}(n)]^{T}$,
for $Z=A,B,C,D,F,G$. The matrices $W_{A}$, $\sigma_{x}$ and $\sigma_{x}'$
appearing in Eqs. (\ref{eq:zz5757-TBeqsDfct}), are given by \begin{subequations}
\label{eq:zz5757-TBeqsDefect2} 
\begin{eqnarray}
W_{A} & = & \left[\begin{array}{cc}
1 & 1\\
e^{2ik_{x}a} & 1
\end{array}\right],\\
\sigma_{x} & = & \left[\begin{array}{cc}
0 & 1\\
1 & 0
\end{array}\right],\\
\sigma_{x}' & = & \left[\begin{array}{cc}
0 & e^{-2ik_{x}a}\\
e^{2ik_{x}a} & 0
\end{array}\right].
\end{eqnarray}
 \end{subequations}

In close analogy with what was done for the case of the \textit{pentagon-only}
defect line, and after some elementary algebraic manipulations, we
can rewrite the above $2\times2$ matrix equations in a more compact
form: \begin{subequations} \label{eq:zz5757-TBeqsDefect3} 
\begin{eqnarray}
\left[\begin{array}{c}
\mathbf{B}(1)\\
\mathbf{A}(1)
\end{array}\right] & = & -\left[\begin{array}{cc}
\frac{\epsilon}{t}\mathbb{I}_{2} & \big(W_{A}\big)^{\dagger}\\
-\mathbb{I}_{2} & 0
\end{array}\right]\left[\begin{array}{c}
\mathbf{A}(1)\\
\mathbf{G}(0)
\end{array}\right],\label{eq:zz5757-BC1}\\
\left[\begin{array}{c}
\mathbf{A}(1)\\
\mathbf{G}(0)
\end{array}\right] & = & -\left[\begin{array}{cc}
\frac{\epsilon}{t}\big(W_{A}\big)^{-1} & \xi_{c}\big(W_{A}\big)^{-1}\\
-\mathbb{I}_{2} & 0
\end{array}\right]\left[\begin{array}{c}
\mathbf{G}(0)\\
\mathbf{F}(0)
\end{array}\right],\label{eq:zz5757-BC2}\\
\left[\begin{array}{c}
\mathbf{G}(0)\\
\mathbf{F}(0)
\end{array}\right] & = & -\left[\begin{array}{cc}
\frac{1}{\xi_{c}}\Big(\frac{\epsilon}{t}\mathbb{I}_{2}+\xi_{a}\sigma_{x}\Big) & \frac{\xi_{b}}{\xi_{c}}\mathbb{I}_{2}\\
-\mathbb{I}_{2} & 0
\end{array}\right]\left[\begin{array}{c}
\mathbf{F}(0)\\
\mathbf{D}(0)
\end{array}\right],\label{eq:zz5757-BC3}\\
\left[\begin{array}{c}
\mathbf{F}(0)\\
\mathbf{D}(0)
\end{array}\right] & = & -\left[\begin{array}{cc}
\frac{1}{\xi_{b}}\Big(\frac{\epsilon}{t}\mathbb{I}_{2}+\xi_{a}\sigma_{x}'\Big) & \frac{\xi_{c}}{\xi_{b}}\mathbb{I}_{2}\\
-\mathbb{I}_{2} & 0
\end{array}\right]\left[\begin{array}{c}
\mathbf{G}(0)\\
\mathbf{C}(0)
\end{array}\right],\label{eq:zz5757-BC4}\\
\left[\begin{array}{c}
\mathbf{D}(0)\\
\mathbf{C}(0)
\end{array}\right] & = & -\left[\begin{array}{cc}
\frac{\epsilon}{t\xi_{c}}\mathbb{I}_{2} & \frac{1}{\xi_{c}}(W_{A})^{\dagger}\\
-\mathbb{I}_{2} & 0
\end{array}\right]\left[\begin{array}{c}
\mathbf{C}(0)\\
\mathbf{B}(-1)
\end{array}\right],\label{eq:zz5757-BC5}\\
\left[\begin{array}{c}
\mathbf{C}(0)\\
\mathbf{B}(-1)
\end{array}\right] & = & -\left[\begin{array}{cc}
\frac{\epsilon}{t}(W_{A})^{-1} & (W_{A})^{-1}\\
-\mathbb{I}_{2} & 0
\end{array}\right]\left[\begin{array}{c}
\mathbf{B}(-1)\\
\mathbf{A}(-1)
\end{array}\right],\label{eq:zz5757-BC6}
\end{eqnarray}
 \end{subequations} where $\mathbb{I}_{2}$ stands for the $2\times2$
unit matrix. Note that Eqs. (\ref{eq:zz5757-TBeqsDefect3}) are now
$4\times4$ matrix equations. In what follows we are going to denote
the matrices in Eqs. (\ref{eq:zz5757-BC1}), (\ref{eq:zz5757-BC2}),
(\ref{eq:zz5757-BC3}), (\ref{eq:zz5757-BC4}), (\ref{eq:zz5757-BC5})
and (\ref{eq:zz5757-BC6}) by $\mathbb{P}_{1}(\epsilon,k_{x})$, $\mathbb{P}_{2}(\epsilon,k_{x})$,
$\mathbf{\mathbb{P}}_{3}(\epsilon,k_{x})$, $\mathbb{P}_{4}(\epsilon,k_{x})$,
$\mathbb{\mathbb{P}}_{5}(\epsilon,k_{x})$ and $\mathbb{\mathbb{P}}_{6}(\epsilon,k_{x})$,
respectively.

The boundary condition matrix equation relating the vectors 
\begin{eqnarray}
\mathbf{L}(1)&=&[A_{1}(k_{x},1),B_{1}(k_{x},1),A_{2}(k_{x},1),B_{2}(k_{x},1)]^{T}
\end{eqnarray}
 and 
\begin{eqnarray}
\mathbf{L}(-1)&=&[A_{1}(k_{x},-1),B_{1}(k_{x},-1),A_{2}(k_{x},-1),B_{2}(k_{x},-1)]^{T} ,
\end{eqnarray}
becomes 
\begin{eqnarray}
\mathbf{L}(1) & = & \mathbb{M}_{5757} . \mathbf{L}(-1).\label{eq:zz5757_TBDC}
\end{eqnarray}
 In the above equation, the boundary condition matrix, $\mathbb{M}_{5757}$,
is a $4\times4$ matrix (in contrast with the case of the \textit{pentagon-only}
defect line) and is given by 
\begin{eqnarray}
\mathbb{M}_{5757} & = & R.\mathbb{P}_{1}.\mathbb{P}_{2}.\mathbb{P}_{3}.\mathbb{P}_{4}.\mathbb{P}_{5}.\mathbb{P}_{6}.R^{T},\label{eq:zz5757_TBDCMat}
\end{eqnarray}
 where, for the sake of simplicity of notation, we have omitted the
dependence of the matrices $\mathbb{P}_{i}$ on $\epsilon$ and on
$k_{x}$; the matrix $R$ in Eq. (\ref{eq:zz5757_TBDC}) makes a basis
change from $\{B_{1},B_{2},A_{1},A_{2}\}$ to $\{A_{1},B_{1},A_{2},B_{2}\}$,
and is defined by 
\begin{eqnarray}
R & = & \left[\begin{array}{cccc}
0 & 0 & 1 & 0\\
1 & 0 & 0 & 0\\
0 & 0 & 0 & 1\\
0 & 1 & 0 & 0
\end{array}\right].\label{eq:Rmatrix}
\end{eqnarray}

It is worth commenting that, if we change to the basis that uncouples
the $+$ and $-$ energy sectors, $\overline{\mathbb{M}}_{5757}(\epsilon,k_{x})=\Lambda\mathbb{M}_{5757}(\epsilon,k_{x})\Lambda^{-1}$
{[}where matrix $\Lambda$ is defined in Eq. (\ref{eq:zzD-BasisChangeMat}){]},
we readily conclude that the boundary condition matrix mixes $+$
and $-$ modes on opposite sides of the defect.

\subsection{The $zz(558)$ defect}

\label{sec:zzD-zz558}

We can employ an entirely analogous procedure to the $zz(558)$ defect.
As before, we can write the Hamiltonian describing such a system,
and from it we write the TB equations at the $zz(558)$ defect {[}see
Fig. \ref{fig:Scheme_zz5757-zz558}(b){]}. These read \begin{subequations}
\label{eq:zz558-TBeqsDfct} 
\begin{eqnarray}
-\frac{\epsilon}{t}\mathbf{A}(1) & = & W_{A}^{\dagger}\mathbf{B}(0)+\mathbf{B}(1),\\
-\frac{\epsilon}{t}\mathbf{B}(0) & = & \xi_{1}\mathbf{D}(0)+W_{A}\mathbf{A}(1),\\
-\frac{\epsilon}{t}\mathbf{D}(0) & = & \xi_{1}\mathbf{B}(0)+\xi_{2}\sigma_{x}'\mathbf{D}(0)+\xi_{1}\mathbf{A}(0),\\
-\frac{\epsilon}{t}\mathbf{A}(0) & = & \xi_{1}\mathbf{D}(0)+W_{A}^{\dagger}\mathbf{B}(-1),\\
-\frac{\epsilon}{t}\mathbf{B}(-1) & = & \mathbf{A}(-1)+W_{A}\mathbf{A}(0),
\end{eqnarray}
 \end{subequations} where, once more we use the notation $\mathbf{Z}(n)=[Z_{1}(n),Z_{2}(n)]^{T}$,
for $Z=A,B,D$.

As before, we can easily rewrite the above $2\times2$ matrix equations
in a more compact form \begin{subequations} \label{eq:zz558-TBeqsDefect3}
\begin{eqnarray}
\left[\begin{array}{c}
\mathbf{B}(1)\\
\mathbf{A}(1)
\end{array}\right] & = & -\left[\begin{array}{cc}
\frac{\epsilon}{t}\mathbb{I}_{2} & \big(W_{A}\big)^{\dagger}\\
-\mathbb{I}_{2} & 0
\end{array}\right]\left[\begin{array}{c}
\mathbf{A}(1)\\
\mathbf{B}(0)
\end{array}\right],\label{eq:zz558-BC1}\\
\left[\begin{array}{c}
\mathbf{A}(1)\\
\mathbf{B}(0)
\end{array}\right] & = & -\left[\begin{array}{cc}
\frac{\epsilon}{t}\big(W_{A}\big)^{-1} & \xi_{1}\big(W_{A}\big)^{-1}\\
-\mathbb{I}_{2} & 0
\end{array}\right]\left[\begin{array}{c}
\mathbf{B}(0)\\
\mathbf{D}(0)
\end{array}\right],\label{eq:zz558-BC2}\\
\left[\begin{array}{c}
\mathbf{B}(0)\\
\mathbf{D}(0)
\end{array}\right] & = & -\left[\begin{array}{cc}
\frac{1}{\xi_{1}}\Big(\frac{\epsilon}{t}\mathbb{I}_{2}+\xi_{2}\sigma_{x}'\Big) & \mathbb{I}_{2}\\
-\mathbb{I}_{2} & 0
\end{array}\right]\left[\begin{array}{c}
\mathbf{D}(0)\\
\mathbf{A}(0)
\end{array}\right],\label{eq:zz558-BC3}\\
\left[\begin{array}{c}
\mathbf{D}(0)\\
\mathbf{A}(0)
\end{array}\right] & = & -\left[\begin{array}{cc}
\frac{\epsilon}{t\xi_{1}}\mathbb{I}_{2} & \frac{1}{\xi_{1}}(W_{A})^{\dagger}\\
-\mathbb{I}_{2} & 0
\end{array}\right]\left[\begin{array}{c}
\mathbf{A}(0)\\
\mathbf{B}(-1)
\end{array}\right],\label{eq:zz558-BC4}\\
\left[\begin{array}{c}
\mathbf{A}(0)\\
\mathbf{B}(-1)
\end{array}\right] & = & -\left[\begin{array}{cc}
\frac{\epsilon}{t}(W_{A})^{-1} & (W_{A})^{-1}\\
-\mathbb{I}_{2} & 0
\end{array}\right]\left[\begin{array}{c}
\mathbf{B}(-1)\\
\mathbf{A}(-1)
\end{array}\right]\,.\label{eq:zz558-BC5}
\end{eqnarray}
 \end{subequations} The $4\times4$ matrices appearing on the right
hand side of Eqs. (\ref{eq:zz558-BC1}), (\ref{eq:zz558-BC2}), (\ref{eq:zz558-BC3}),
(\ref{eq:zz558-BC4}) and (\ref{eq:zz558-BC5}) will be denoted by
$\mathbb{Q}_{1}(\epsilon,k_{x})$, $\mathbb{\mathbb{Q}}_{2}(\epsilon,k_{x})$,
$\mathbb{Q}_{3}(\epsilon,k_{x})$, $\mathbb{Q}_{4}(\epsilon,k_{x})$
and $\mathbb{Q}_{5}(\epsilon,k_{x})$, respectively.

The boundary condition matrix equation relating $\mathbf{L}(1)$ and
$\mathbf{L}(-1)$ is of the same form as in Eq.~(\ref{eq:zz5757_TBDC})
\begin{eqnarray}
\mathbf{L}(1) & = & \mathbb{M}_{558}.\mathbf{L}(-1),\label{eq:zz558_TBDC}
\end{eqnarray}
 where, the boundary condition matrix, $\mathbb{M}_{558}$, is a $4\times4$
matrix now given by 
\begin{eqnarray}
\mathbb{M}_{558} & = & R.\mathbb{Q}_{1}.\mathbb{Q}_{2}.\mathbb{Q}_{3}.\mathbb{Q}_{4}.\mathbb{Q}_{5}.R^{T},\label{eq:zz558_TBDCMat}
\end{eqnarray}
 where again, for the sake of simplicity of notation, we have omitted
the dependence of the matrices $\mathbb{Q}_{i}$ on $\epsilon$ and
on $k_{x}$.

Once again, in the basis that uncouples the $+$ and the $-$ modes,
$\overline{\mathbb{M}}_{558}(\epsilon,k_{x})=\Lambda\mathbb{M}_{558}\Lambda^{-1}$
{[}where matrix $\Lambda$ is defined in Eq. (\ref{eq:zzD-BasisChangeMat}){]},
one readily sees that the boundary condition matrix also mixes the modes
of the $+$ and the $-$ energy sectors of opposite sides of the defect.

\subsection{Computing the scattering coefficients}

\label{sec:zzD-ScattCoeffs}

The scattering problem from the $zz(5757)$ {[}and the $zz(558)${]}
defect line can now be solved completely using the boundary condition
at the defect, Eqs. (\ref{eq:zz5757_TBDCMat}) {[}Eq. (\ref{eq:zz558_TBDCMat}){]},
and the modes allowed in the bulk {[}obtained from the transfer matrix,
Eq. (\ref{eq:zzD-doubledUCpassageM})-(\ref{eq:zzD-doubleUCpassageMats}){]}.

Let us suppose that we are working at positive energy and around the
$\nu=+1$ Dirac point {[}$\mathbf{K}_{+}=\pi/3(1/a,-\sqrt{3}/a)${]}.
Moreover, we suppose that the energy is small, so that the $+$ modes
are evanescent. We then consider the scattering process in which we
have one incoming mode from $n=-\infty$ associated with the $-$
energy sector. Due to the presence of the defect at $n=0$, there
will be a reflected as well as a transmitted mode. Besides, in accordance
with what was said above, there will be two evanescent modes associated
with the $+$ energy sector.

We choose to denote $\vert\Psi_{>}^{(-)}\rangle$ ($\vert\Psi_{<}^{(-)}\rangle$)
as the right (left) moving mode associated with the $-$ energy sector
of the transfer matrix {[}see Eqs. (\ref{eq:zzD-PassMat1})-(\ref{eq:zzD-doubleUCpassageMats}){]}
when we are both at $\epsilon>0$ and around the $\mathbf{K}_{+}$
Dirac point; the corresponding eigenvalues are denoted by $\lambda_{>}^{(-)}$
($\lambda_{<}^{(-)})$. Similarly, we denote by $\vert\Psi_{>}^{(+)}\rangle$
($\vert\Psi_{<}^{(+)}\rangle$) the $+$ energy sector's
transfer matrix mode {[}see Eq. (\ref{eq:zzD-PassMat1})-(\ref{eq:zzD-doubleUCpassageMats}){]},
which decreases (increases) in the direction of $\mathbf{u}_{2}$.
This mode's corresponding eigenvalue is going to be denoted by $\lambda_{>}^{(+)}$
($\lambda_{<}^{(+)}$.) We can thus write the wave function on each
side of the defect as \begin{subequations} \label{eq:zzD-WaveFunctions}
  \begin{eqnarray}
    \mathbf{L}(n<0) & = & \big(\lambda_{>}^{(-)}\big)^{n+1} \vert \Psi_{>}^{(-)} \rangle 
    +\rho_{-}\big(\lambda_{<}^{(-)}\big)^{n+1} \vert \Psi_{<}^{(-)} \rangle \nonumber \\ 
    & & + \rho_{+}\big(\lambda_{<}^{(+)}\big)^{n+1} \vert \Psi_{<}^{(+)} \rangle ,\\
    \mathbf{L}(n>0) & = & \tau_{-}\big(\lambda_{>}^{(-)}\big)^{n-1} \vert \Psi_{>}^{(-)} \rangle 
    + \tau_{+}\big(\lambda_{>}^{(+)}\big)^{n-1}\vert \Psi_{>}^{(+)} \rangle ,
  \end{eqnarray}
 \end{subequations} where $\rho_{-}$ and $\tau_{-}$ ($\rho_{+}$
and $\tau_{+}$) are the reflection and transmission scattering amplitudes
of the $-$ ($+$) modes, respectively.

As before, for different choices of $\epsilon$ and $k_{x}$ the incoming,
reflected, transmitted and evanescent modes must be chosen accordingly
with the direction of propagation and/or direction of increase of
the eigenmodes of the transfer matrix (see Fig. \ref{fig:CurrModsDoubled}).

In order to determine the scattering coefficients $\rho_{\pm}$ and
$\tau_{\pm}$, and in accordance with Section \ref{sec:GenDescrp},
we just need to substitute Eqs. (\ref{eq:zzD-WaveFunctions}) in the
boundary condition expression arising from the $zz(5757)$ {[}$zz(558)${]}
defect, Eq. (\ref{eq:zz5757_TBDC}) {[}Eq. (\ref{eq:zz558_TBDC}){]}.
to obtain the equivalent of Eq.~(\ref{eq:BCGen_PropBasis}) for these
cases, 
\begin{eqnarray}
\left[\begin{array}{c}
\tau_{+}\\
0\\
\tau_{-}\\
0
\end{array}\right]_{n=+1} & = & U^{-1}\mathbb{M}U\left[\begin{array}{c}
0\\
\rho_{+}\\
1\\
\rho_{-}
\end{array}\right]_{n=-1},\label{eq:zzD-TB_BC2}
\end{eqnarray}
where we have omitted the subscripts identifying the boundary condition
matrix {[}either $\mathbb{M}{}_{5757}$ for the $zz(5757)$ defect,
or $\mathbb{M}_{558}$ for the $zz(558)${]}. In the above equation,
the matrix $U$ is the matrix mediating the basis transformation from
the basis $\{A_{+},B_{+},A_{-},B_{-}\}$ to the proper basis of the
transfer matrix. As stated in Section \ref{sec:GenDescrp}, each column
of $U$ is one of the eigenmodes of the transfer matrix $U=\left[ \vert \Psi_{>}^{(+)}\rangle ,\vert \Psi_{<}^{(+)}\rangle ,\vert \Psi_{>}^{(-)}\rangle ,\vert \Psi_{<}^{(-)}\rangle \right]$,
written in the $\{A_{+},B_{+},A_{-},B_{-}\}$ basis. 

Solving the above linear system of four equations can be readily accomplished
with a computer algebra system. Its solution determines completely
the scattering amplitudes of the problem.

\subsection{The transmittance}

\label{sec:zzD-Transmtt}

As before, we are interested in computing the transmittance for relatively
low-energies, which is equivalent to say that we want to compute the
transmittance around the Dirac points. Consequently, the $+$ modes
will be evanescent while the $-$ modes will be propagating. Therefore
the transmittance is going to be given by $T=\vert\tau_{-}\vert^{2}$.

Similarly to what was done for the transmittance in the \textit{pentagon-only}
defect case, here, and for the sake of comparison between scattering
processes occurring at positive and negative energy, we will plot
the transmittance against $\theta'=\theta$ when $\epsilon>0$, plotting
it against $\theta'=-\theta$ when $\epsilon<0$.

\begin{figure}[htp!]
  \centering
  \includegraphics[width=0.48\textwidth]{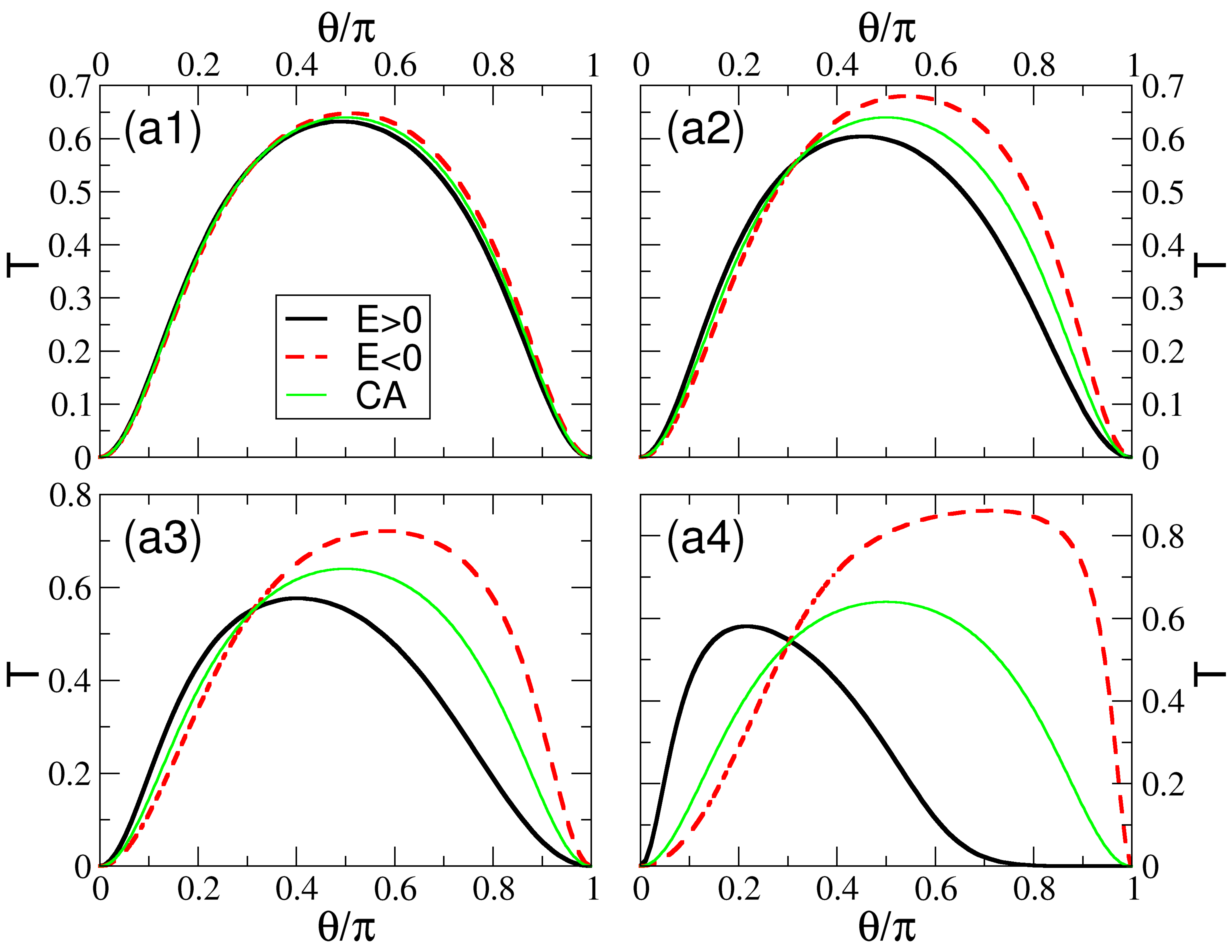}
  \includegraphics[width=0.48\textwidth]{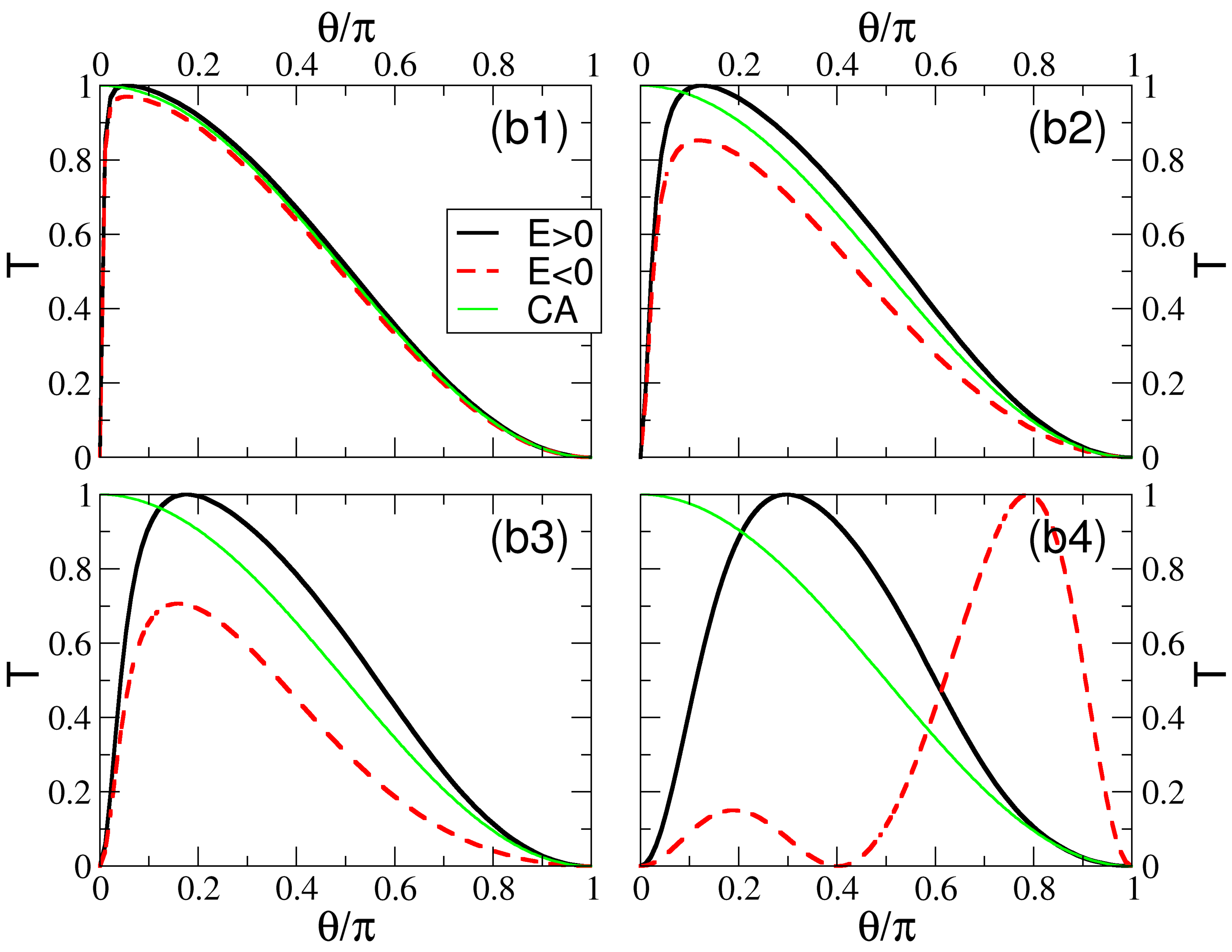}
  \caption{(Color online) Panels (a): Transmittance through the
    $zz(5757)$ defect line as function of the angle $\theta$ (when
    $\epsilon>0$ and $-\theta$ when $\epsilon<0$) between $\mathbf{q}$
    and the barrier. The hopping parameters at the defect have the
    value $\xi_{a}=\xi_{b}=\xi_{c}=1$.  Panels (b): Transmittance
    through the $zz(558)$ defect line as function of the angle
    $\theta$ (when $\epsilon>0$ and $-\theta$ when $\epsilon<0$)
    between $\mathbf{q}$ and the barrier. The hopping parameters at
    the defect are $\xi_{1}=\xi_{2}=1$. In each group of panels, the
    transmittance is plotted for several different energy
    modulus. These are: (a1) and (b1) $\vert\epsilon/t\vert=0.01$;
    (a2) and (b2) $\vert\epsilon/t\vert=0.05$; (a3) and (b3)
    $\vert\epsilon/t\vert=0.1$; (a4) and (b4)
    $\vert\epsilon/t\vert=0.3$. Positive energies are represented by
    the black full lines, while the negative ones are represented by
    the dashed red lines. The green full lines stand for the continuum
    low-energy result (that is energy-independent) as obtained in
    Ref. \cite{Rodrigues_PRB:2012}. Only the curves associated with
    the Dirac point $\mathbf{K}_{+}$ are represented.  Those for the
    Dirac point $\mathbf{K}_{-}$ are obtained from the former by a
    reflection of these over the axis $\theta=\pi/2$.}
  \label{fig:zzD-theta_pos_neg_E}
\end{figure}

In Fig. \ref{fig:zzD-theta_pos_neg_E} we represent the transmittance
$T=\vert\tau_{-}\vert^{2}$ {[}associated both with the $zz(5757)$
and the $zz(558)$ defect lines{]} as function of the angle $\theta$
(when $\epsilon>0$ and $-\theta$ when $\epsilon<0$), made between
$\mathbf{q}$ and the defect line. In the several panels of Fig. \ref{fig:zzD-theta_pos_neg_E},
we present the transmittance curves for different values of the energy
{[}the ones referring to the $zz(5757)$ defect are those in panels
(a), while the ones referring to the $zz(558)$ defect are those in
panels (b){]}. The curves in the Fig. \ref{fig:zzD-theta_pos_neg_E}
refer to the Dirac point $\mathbf{K}_{+}$. Those referring to the
Dirac point $\mathbf{K}_{-}$, are mirror symmetric relatively to
the line $\theta=\pi/2$ to the ones presented in the figure. In order
to obtain the transmittance curves of Fig. \ref{fig:zzD-theta_pos_neg_E},
all hopping renormalizations were set equal to one in both defects:
$\xi_{a}=\xi_{b}=\xi_{c}=1$ and $\xi_{1}=\xi_{2}=1$.

Again, the transmittance profile arising from each of these defect lines is controlled by the values
of the hopping parameters at the defect line: $\xi_{1}$ and $\xi_{2}$ in the case of $zz(558)$ defect line;
$\xi_{a}$, $\xi_{b}$ and $\xi_{c}$ in the case of the $zz(5757)$ defect line. Modifications of these parameters
result in very different transmittance behaviors.

Here as in the {\it pentagon-only} defect case, the analysis of the TB expressions giving the transmission 
amplitude, Eq. (\ref{eq:zzD-TB_BC2}), is of little help concerning physical intuition over the presence of 
perfect transmittance angles for certain ranges of values for the hopping parameters. As before, this feature 
is more easily interpreted in the continuum low-energy description of these systems [verify the good accordance 
between the TB and the low-energy description of these systems in Fig. \ref{fig:zzD-theta_pos_neg_E}(a1) and 
(b1)].

As stated in \ref{app:CAtransm}, in the continuum low-energy limit, both the $zz(558)$ and the $zz(5757)$ defect 
lines can be viewed as a strip with a general potential which has $V_{s} \neq 0 \neq V_{x}$ and $V_{y} = 0 = V_{z}$
[see Eqs. (\ref{eq:GenPotential-558}) and Eqs. (\ref{eq:GenPotential-5757})]. We argued, at the end of section 
\ref{sec:JS-transmtt}, concerning the {\it pentagon-only} defect, that perfect transmission occurs when the spin 
directions of the incident and transmitted waves coincide. This interpretation can be carried over the $zz(558)$ 
and $zz(5757)$ defects.

In the case of a $zz(558)$ defect line with hopping parameters $\xi_{1} = \xi_{2} = 1$ [see Figs. 
\ref{fig:zzD-theta_pos_neg_E}(b)], the angle of perfect transmission is $\widetilde{\alpha}=\arccos(\xi_{2}
/\xi_{1}^{2})$ (see \ref{app:CAtransm}); for hopping parameters $\xi_{1}=\xi_{2}=1$ we obtain $\widetilde{\alpha}=0$, 
in accordance with what one can see in Fig. \ref{fig:zzD-theta_pos_neg_E}(b1). The
fermions incident on the defect line with an angle $\theta \approx 0$ [and thus, a spin oriented along 
this direction] have their spin oriented in the same direction as those propagating inside the strip.
Therefore, they will be perfectly transmitted, and all the others will be partly reflected.

Similarly, for a $zz(5757)$ defect line with $\xi_{a} = \xi_{b} = \xi_{c} = 1$, there is no angle with 
perfect transmission, since $\bar{\alpha} = \arccos [- (a + c)/ 2 d]$ [see \ref{app:CAtransm}] is not a real 
number when $\xi_{a} = \xi_{b} = \xi_{c} = 1$ [in accordance with what one can see in Fig. 
\ref{fig:zzD-theta_pos_neg_E}(a1)]. Inside the strip, the spin of the fermions has a component in 
the $z$-plane, and thus will never be totally aligned with the spin of any fermion incident on the 
defect line.

\section{Conclusions}

\label{sec:conclusion}

In this paper we have presented a procedure to work out the electronic
scattering either from a defect in a one-dimensional crystal, or from
a periodic extended defect in two-dimensional crystals, which can
be reduced to a quasi-one-dimensional tight-binding model. The formalism 
developed only uses simple tight-binding concepts, which reduce the scattering
problem to a set of matrix manipulations. These can be easily worked out by 
any computational algebraic calculator.

We have illustrated the presented procedure in the context of the first neighbor
tight-binding model of graphene when a defect line is oriented along
the zigzag direction in an otherwise perfect crystal. Three distinct
defects were studied: the \textit{pentagon-only}, the $zz(5757)$,
and the $zz(558)$ defect lines.

The latter two defect lines forced a duplication of the unit cell
along the defect direction (relatively to the case of \textit{pentagon-only}
defect line). Such duplication renders the mathematical treatment
of the problem more complex, by introducing an additional pair of
scattering modes. These latter modes happen to be propagating at high
energy around the Dirac points. Thus, for low-energy scattering processes
they will be evanescent states.

All the three defect lines studied have different behaviors that strongly
depend on the particular values of the hopping parameters at the defect
lines. Moreover, in a companion paper \cite{Rodrigues_PRB:2012}
we treat these same problems in the continuum low-energy limit using
the Dirac equation. We have shown that we can recover the low energy results
from the tight-binding approach, as it should be the case.

Noteworthy is the fact that the present procedure can be used to treat
more realistic extended periodic defects, that do not need to be linear.
In principle, such kind of extended defects will force a considerable
increase in the size of the unit cell, thus introducing a larger number
of scattering modes. This, of course, renders the problem a more difficult
one.

\appendix


\section{Left and right eigenvectors of the transfer matrix $\mathbb{T}$}

\label{app:DualBasis}

The right eigenvectors of a non-hermitian matrix $\mathbb{T}$, are
defined by the equation 
\begin{eqnarray}
\mathbb{T} & \left|\psi_{i}\right\rangle = & \lambda_{i}\left|\psi_{i}\right\rangle ,\label{eq:RightEig}
\end{eqnarray}
 while the left ones, are defined by 
\begin{eqnarray}
\langle \widetilde{\psi}_{i} \vert \mathbb{T} & = & \lambda_{i} \langle \widetilde{\psi}_{i}\vert . \label{eq:LeftEig}
\end{eqnarray}

It is straightforward to demonstrate that, apart from a normalization
constant, $C_{i}=\sqrt{\langle \widetilde{\psi}_{i}\vert \psi_{i} \rangle}$,
they form a dual basis. Explicitly, 
\begin{eqnarray}
\langle \widetilde{\psi}_{j}\vert\mathbb{T}\vert\psi_{i}\rangle  & = & \lambda_{j}\langle \widetilde{\psi}_{j}\vert \psi_{i} \rangle =\lambda_{i}\langle \widetilde{\psi}_{j}\vert \psi_{i} \rangle ,\label{eq:DualBasisProof}
\end{eqnarray}
and then, if $\lambda_{i}\neq\lambda_{j}$, we must have $\langle \widetilde{\psi}_{j}\vert\psi_{i}\rangle =0$.
If we define $\vert\varphi_{i}\rangle \equiv\vert\psi_{i}\rangle /C_{i}$
and $\vert\widetilde{\varphi}_{i}\rangle \equiv\vert\widetilde{\psi}_{i}\rangle /C_{i}$,
we can thus write 
\begin{eqnarray}
\langle \widetilde{\varphi}_{j}\vert \varphi_{i}\rangle  & = & \delta_{ij}.\label{eq:DualBasis}
\end{eqnarray}

Let us now show that the left eigenvectors of a non-hermitian matrix,
$\mathbb{T}$, are the right eigenvectors of its transpose, $\mathbb{T}^{T}$.
If we take the hermitian conjugate of Eq. (\ref{eq:LeftEig}), we
obtain 
\begin{eqnarray}
\mathbb{T}^{\dagger}\vert\widetilde{\psi}_{i}\rangle  & = & \lambda_{i}^{*}\vert\widetilde{\psi}_{i}\rangle .\label{eq:LeftEigDagger}
\end{eqnarray}
 Spelling out this equation in components, and taking the complex
conjugate, we obtain 
\begin{eqnarray}
\big(\mathbb{T}^{T}\big)_{\alpha\beta}\big(\vert\widetilde{\psi}_{i}\rangle \big)_{\beta}^{*} & = & \lambda_{i}\big(\vert\widetilde{\psi}_{i}\rangle \big)_{\alpha}^{*},\label{eq:LeftEigCC}
\end{eqnarray}
which gives us the components of the row vector $\langle \widetilde{\psi}_{j}\vert_{\alpha}=\big(\vert\widetilde{\psi}_{i}\rangle \big)_{\alpha}^{*}$.

\section{The conserved current along the $1$D chain}

\label{app:ConsvCurr}

In this appendix we are going to compute the conserved current along
a quasi-one-dimensional chain. We start by doing it for a general
case (using the formalism of Section \ref{sec:GenDescrp}). Later,
we write the conserved current for the one-dimensional chains arising
from the Fourier transformation of both the graphene layer with a
\textit{pentagon-only} defect line, and that with a $zz(5757)$ {[}or
a $zz(558)${]} defect line.

The bulk TB equations of the general quasi-one-dimensional chain,
Eq. (\ref{eq:GenTBeqs}), can be rewritten in the following manner
\begin{eqnarray}
\epsilon\mathbf{c}(n) & = & V_{L}^{\dagger}\mathbf{c}(n-1)+H_{r}\mathbf{c}(n)+V_{L}\mathbf{c}(n+1).\label{eq:GenTBeqsExplct}
\end{eqnarray}
 where we have used $V_{R}=V_{L}^{\dagger}$ (Hamiltonian hermiticity). The time-dependent
counterpart of this equation can be written as 
\begin{eqnarray}
i\hbar\frac{\partial}{\partial t}\mathbf{c}(n,t) & = & V_{L}^{\dagger}\mathbf{c}(n-1,t)+H_{r}\mathbf{c}(n,t)+V_{L}\mathbf{c}(n+1,t).\label{eq:GenTBeqsTime}
\end{eqnarray}
 The adjoint of Eq. (\ref{eq:GenTBeqsTime}) reads 
\begin{eqnarray}
-i\hbar\frac{\partial}{\partial t}\mathbf{c}^{\dagger}(n,t) & = & \mathbf{c}^{\dagger}(n-1,t)V_{L}+\mathbf{c}^{\dagger}(n,t)H_{r}+\mathbf{c}^{\dagger}(n+1,t)V_{L}^{\dagger},\label{eq:GenTBeqsTimeAdj}
\end{eqnarray}
 where $H_{r}^{\dagger}=H_{r}$ from the TB Hamiltonian hermiticity
property. Given this, we can write 
\begin{eqnarray}
\frac{\partial}{\partial t}\mathbf{c}^{\dagger}(n,t)\mathbf{c}(n,t) & = & -\Big[\mathcal{J}(n,t)-\mathcal{J}(n-1,t)\Big],\label{eq:GenTBeqsTimeN}
\end{eqnarray}
 where 
\begin{eqnarray}
\mathcal{J}(n,t) & = & \frac{i}{\hbar}\Big[\mathbf{c}^{\dagger}(n,t)V_{L}\mathbf{c}(n+1,t)-\mathbf{c}^{\dagger}(n+1,t)V_{L}^{\dagger}\mathbf{c}(n,t)\Big],\label{eq:GenTBConsCurr}
\end{eqnarray}
 can be interpreted as the particle current along the one-dimensional
chain flowing from position $n$ to position $n+1$. If we have a
stationary state of an $AB$-like chain, we can use the transfer matrix
relation {[}see Eq. (\ref{eq:TransfMat-AB}){]} and write the conserved
current as 
\begin{eqnarray}
\mathcal{J}(n) & = & \mathbf{c}^{\dagger}(n)\bigg[\frac{i}{\hbar}\Big(V_{L}\mathbb{T}-\mathbb{T}^{\dagger}V_{L}^{\dagger}\Big)\bigg]\mathbf{c}(n).\label{eq:GenTBConsCurr2}
\end{eqnarray}
 We can further rewrite the conserved current expression using the
fact that Bloch modes are the eigenvectors of the transfer matrix,
$\mathbb{T}\mathbf{\psi}_{\lambda}=\lambda\mathbf{\psi}_{\lambda}$.
Therefore, for a Bloch solution labeled by $\lambda$, whose eigenvector
is $\mathbf{\psi}_{\lambda}$, the conserved current reads 
\begin{eqnarray}
\mathcal{J}_{\lambda} & = & \mathbf{\psi}_{\lambda}^{\dagger}\bigg[\frac{i}{\hbar}\Big(\lambda V_{L}-\lambda^{*}V_{L}^{\dagger}\Big)\bigg]\mathbf{\psi}_{\lambda}.\label{eq:GenTBConsCurr3}
\end{eqnarray}

For the case of the \textit{pentagon-only} defect line it can be inferred
from Eqs. (\ref{eq:TBbulkEqs}) that the matrix $V_{L}$ reads 
\begin{eqnarray}
V_{L} & = & -t\left[\begin{array}{cc}
0 & 0\\
1+e^{ik_{x}a} & 0
\end{array}\right],\label{eq:VL-grph}
\end{eqnarray}
 while the transfer matrix is given by Eq. (\ref{eq:BulkPassageM}).
As a consequence, we can easily conclude that the conserved current
operator along the one-dimensional chain (originated from the Fourier
transformation of pristine graphene along the direction of $\mathbf{u}_{1}$),
is given by 
\begin{eqnarray}
\hat{\mathbf{\mathcal{J}}} & = & \frac{i}{\hbar}\Big(V_{L}\mathbb{T}-\mathbb{T}^{\dagger}V_{L}^{\dagger}\Big)=\frac{t}{\hbar}\sigma_{y},\label{eq:zz_ConsCurr}
\end{eqnarray}
 where $\sigma_{y}$ is the $y$-Pauli matrix. As a consequence, the
conserved current reads 
\begin{eqnarray}
\mathcal{J}_{\lambda} & = & \frac{t}{\hbar}\mathbf{\psi}_{\lambda}^{\dagger}\sigma_{y}\mathbf{\psi}_{\lambda}.\label{eq:zz_ConsCurr2}
\end{eqnarray}

Note that in the context of the two-dimensional graphene lattice,
the above conserved current is a current directed along $\mathbf{u}_{2}$.
Therefore, the current along the $y$-direction is given by $\mathcal{J}_{y}(\lambda)=\mathcal{J}_{\lambda}\sqrt{3}/2$,
which is in accordance with the low-energy result.
\begin{figure}[htp!]
  \centering
  \includegraphics[width=0.95\textwidth]{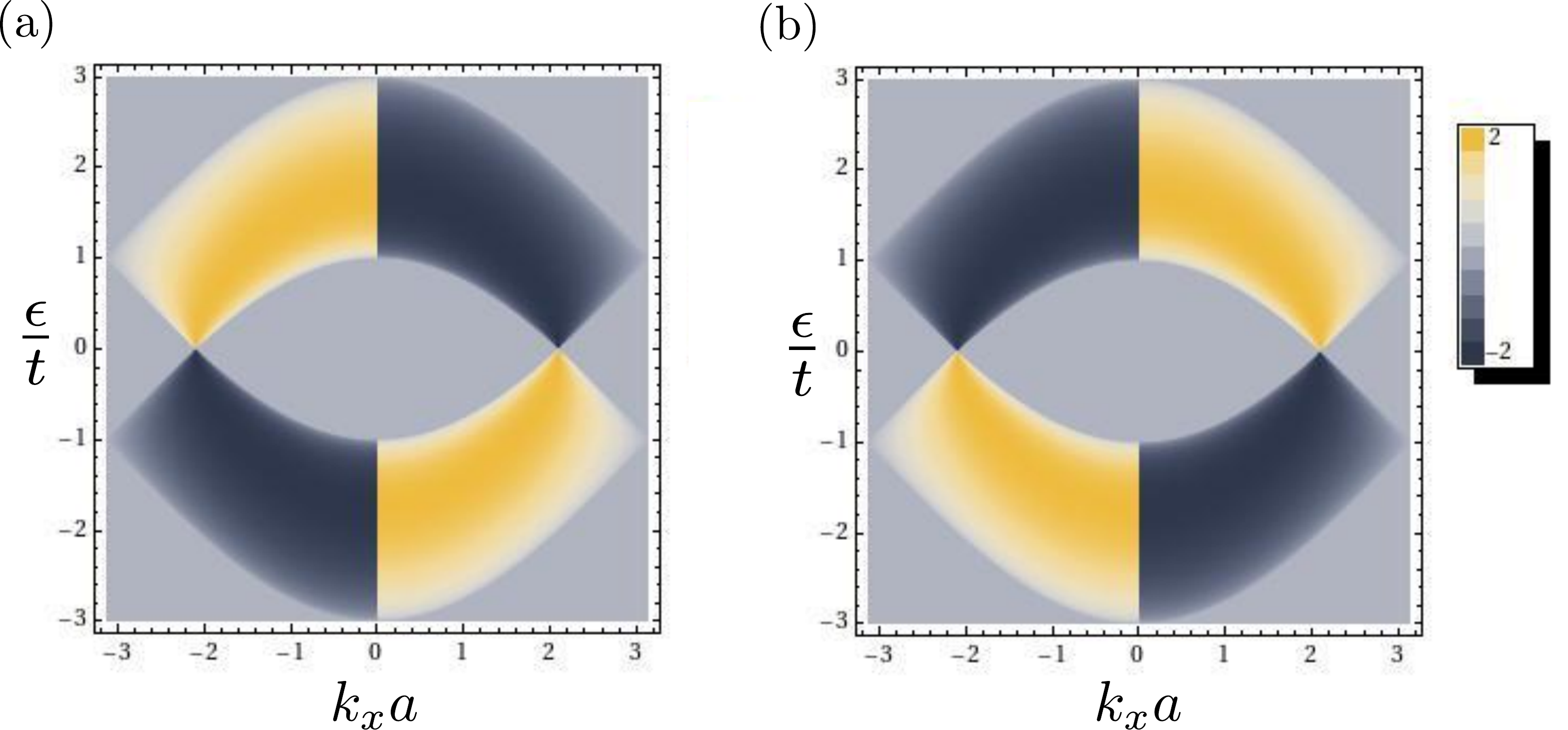}
  \caption{(Color online) Panels (a) and (b) are density plots of the
    current associated with the modes of the transfer matrix of the
    pristine graphene lattice. In accordance with the definition of
    sub-Section \ref{sec:JS-SctCoeffs}, panel (a) gives the current
    associated with the mode $\vert\Psi_{>}\rangle$, while panel (b)
    gives the current associated with the mode
    $\vert\Psi_{<}\rangle$.}
  \label{fig:CurrModsPristine}
\end{figure}

In Fig. \ref{fig:CurrModsPristine} we present a density plot of the
current associated with each one of the modes of the transfer matrix
of pristine graphene.

Similarly, for a one-dimensional chain obtained from the Fourier transform
of graphene with a doubled unit cell in the direction $\mathbf{u}_{1}$
{[}which happens in the cases of the $zz(5757)$ and $zz(558)$ defect
lines{]}, we can infer from the bulk equations, Eqs. (\ref{eq:zzD-TBbulkEqs}),
that the matrix $V_{L}$ reads 
\begin{eqnarray}
V_{L} & = & -t\left[\begin{array}{cccc}
0 & 0 & 0 & 0\\
1 & 0 & 1 & 0\\
0 & 0 & 0 & 0\\
e^{i2k_{x}a} & 0 & 1 & 0
\end{array}\right],\label{eq:VL-Dgrph}
\end{eqnarray}
 while the transfer matrix for this quasi-one-dimensional chain is
given by Eq. (\ref{eq:zzD-PassMat1}). As a consequence, we can readily
conclude that the conserved current operator along this chain, is
given by 
\begin{eqnarray}
\hat{\mathbf{\mathcal{J}}} & = & \frac{i}{\hbar}\Big(V_{L}\mathbb{T}-\mathbb{T}^{\dagger}V_{L}^{\dagger}\Big)=\frac{t}{\hbar}\left[\begin{array}{cc}
\sigma_{y} & 0_{2}\\
0_{2} & \sigma_{y}
\end{array}\right],\label{eq:zzD_ConsCurr}
\end{eqnarray}
 where $\sigma_{y}$ stands for the $y$-Pauli matrix, while $0_{2}$
represents a $2\times2$ zero matrix. Note that the appearance of
two copies of pristine graphene conserved current was to be expected because,
as was already referred, the folding of the FBZ due to the duplication
of the unit cell along the direction of $\mathbf{u}_{1}$ brings an
additional pair of modes into the folded Brillouin zone. Both the
modes associated with the $+$ and the $-$ energy sector were already
known to have a current operator given by Eq. (\ref{eq:zz_ConsCurr}).
\begin{figure}[htp!]
  \centering
  \includegraphics[width=0.85\textwidth]{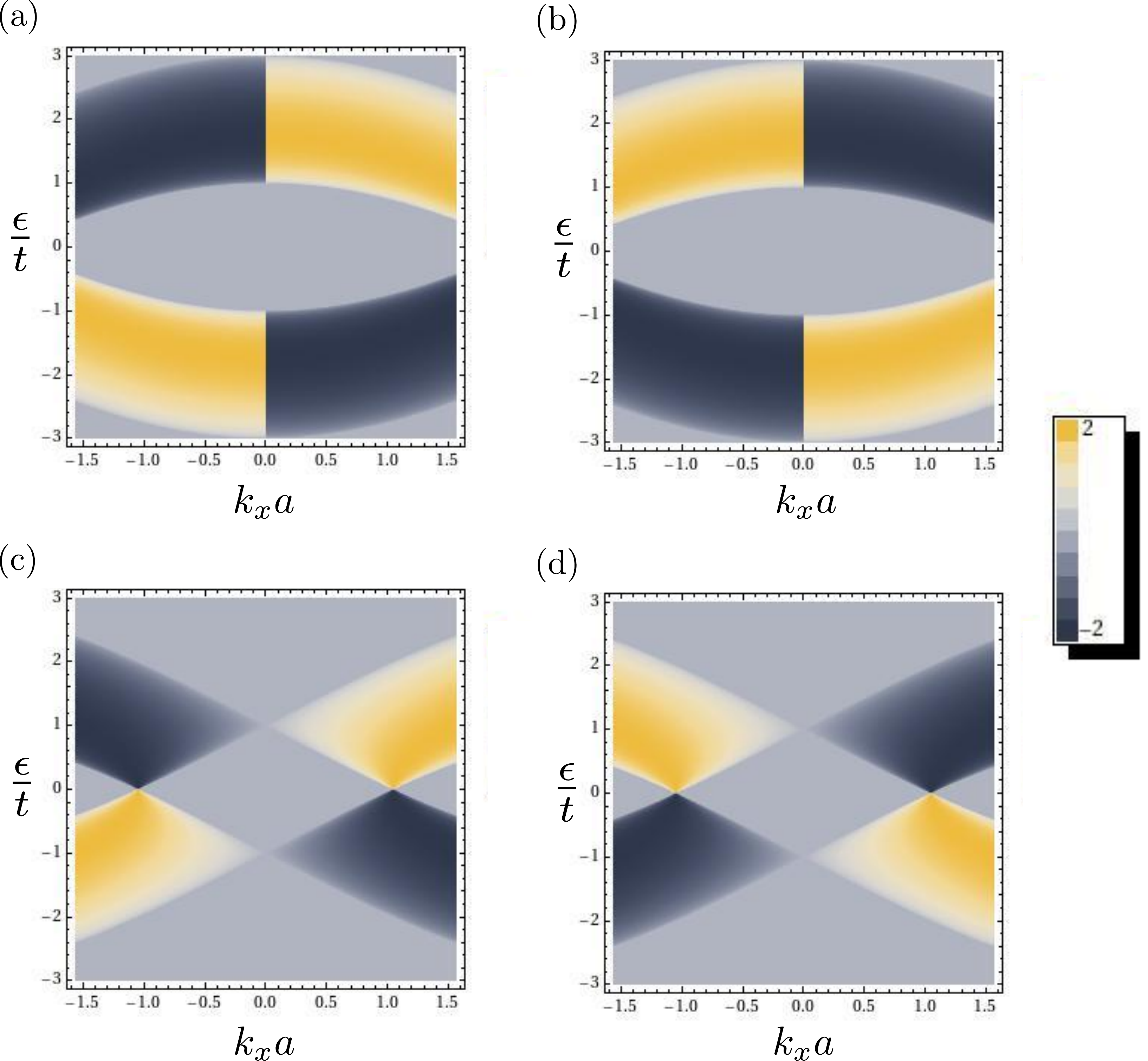}
  \caption{(Color online) Panels (a)-(d) are density plots of the
    current associated with the modes of the transfer matrix of the
    graphene lattice doubled along the $\mathbf{u}_{1}$ direction. In
    accordance with the definition of sub-Section
    \ref{sec:zzD-ScattCoeffs}, panel (a) {[}(b){]} gives the current
    associated with the $(+$) modes, while panel (c) {[}(d){]} gives
    the current associated with the $(-)$ modes.}
  \label{fig:CurrModsDoubled}
\end{figure}

In Fig. \ref{fig:CurrModsDoubled} we present a density plot of the
particle current associated with each one of the modes of the transfer
matrix of graphene with a doubled unit cell along the direction of
$\mathbf{u}_{1}$.


\section{The basis uncoupling the high and low-energy modes}

\label{app:BasisUncp}

As we have argued in the main text it is, in some circumstances, required
to work with a doubled unit cell. That is, for instance, the case
when treating the $zz(5757)$ and the $zz(558)$ defect lines in graphene.

With either of these two defect lines present, $\mathbf{u}_{1}$ is
no longer a lattice translation. To accommodate this, it is convenient
to describe pristine graphene with a unit lattice defined by the vectors
$2\mathbf{u}_{1}$ and $\mathbf{u}_{2}$, instead of the usual choice,
$\mathbf{u}_{1}$ and $\mathbf{u}_{2}$. The corresponding reciprocal
lattice vectors will then be $\mathbf{v}_{1}/2$ and $\mathbf{v}_{2}$,
instead of those associated to the FBZ of pristine graphene, $\mathbf{v}_{1}=2\pi/a(1,1/\sqrt{3})$
and $\mathbf{v}_{2}=2\pi/a(0,2/\sqrt{3})$. Therefore, the doubling
of the unit cell will have as its main consequence the folding of
the FBZ along the direction of $\mathbf{v}_{1}$ {[}compare both panels
of Fig. \ref{fig:FBZs}{]}.

There will thus be twice as many atoms in the doubled unit cell as
those contained in the pristine graphene one {[}two atoms of sub-lattice
$A$ ($A_{1},\, A_{2}$) and two atoms of sub-lattice $B$ ($B_{1},\, B_{2}$){]}.
At the same time, there will be twice as many energy bands in the
folded FBZ as those present in the pristine graphene one. Moreover,
the Dirac points will now be located at $\mathbf{K}_{\nu}=\nu\pi/3(1/a,-\sqrt{3}/a)$
(see Fig. \ref{fig:FBZs}). Near the \textit{new } Dirac points, two
of the four energy bands present in the folded FBZ are the two low-energy
Dirac cones, while the other two bands are the high-energy bands \cite{Rodrigues_PRB:2012,Rodrigues_PRB:2011}.
In what follows, we identify the two bands of low-energy near the
Dirac points by $-$, while the high-energy ones are identified by
$+$.

The bulk physics of graphene with a doubled unit cell must be exactly
the same as that of pristine graphene. In fact, the Bloch vectors $\mathbf{q}$
and $\mathbf{q}+\mathbf{v}_{1}/2$ identifying two different Bloch
wave solutions in the \emph{unfolded} Brillouin zone of pristine graphene
correspond to a single Bloch vector $\mathbf{q}$ in the folded zone.
As a consequence, in the case of graphene with a doubled unit cell
it is natural to expect that we can find a basis in which we can uncouple
the physics associated with each one of the two Bloch solutions of
the \textit{unfolded} system: that identified by the Bloch vector
$\mathbf{q}$ and that identified by the Bloch vector $\mathbf{q}+\mathbf{v}_{1}/2$.

Let us then put ourselves at the wave vector $\mathbf{k}=\mathbf{q}$,
where $\mathbf{q}$ is close to the Dirac point of the folded Brillouin
zone. Bloch theorem in the \textit{unfolded} pristine graphene allows
us to write the following relations between amplitudes of the doubled
unit cell: \begin{subequations} \label{eq:BulkRels1} 
\begin{eqnarray}
A_{2} & = & e^{i\mathbf{q}\cdot\mathbf{u}_{1}}A_{1}=e^{iq_{x}a}A_{1},\\
B_{2} & = & e^{i\mathbf{q}\cdot\mathbf{u}_{1}}B_{1}=e^{iq_{x}a}B_{1}.
\end{eqnarray}
 \end{subequations}

Similarly, for the wave vector $\mathbf{k}=\mathbf{q}+\mathbf{v_{1}}/2$,
we have the relations \begin{subequations} \label{eq:BulkRels2}
\begin{eqnarray}
A_{2} & = & e^{i\big(\mathbf{q}+\frac{\mathbf{v_{1}}}{2}\big)\cdot\mathbf{u}_{1}}A_{1}=-e^{iq_{x}a}A_{1},\\
B_{2} & = & e^{i\big(\mathbf{q}+\frac{\mathbf{v_{1}}}{2}\big)\cdot\mathbf{u}_{1}}B_{1}=-e^{iq_{x}a}B_{1}.
\end{eqnarray}
 \end{subequations} We note that the vector $\mathbf{q}+\mathbf{v_{1}}/2$
can be close to the Dirac point of the unfolded zone. In fact, if
we choose $\mathbf{q}=\bm{K}_{D}=(\pi/3a,-\pi/\sqrt{3}a)$, the Dirac
point of the folded zone, $\mathbf{q}+\mathbf{v}_{1}/2=(4\pi/3a,0)$,
the Dirac point of the unfolded zone. Then, in the unfolded zone,
the state with $\mathbf{k}=\mathbf{q}$ corresponds to a high energy
state (+), whereas the state $\mathbf{k}=\mathbf{q}+\mathbf{v_{1}}/2$
corresponds to a low energy state (-). This analysis motivates the
past definitions of high and low energy modes.

As said above, we want to construct a basis that uncouples the modes
originating at the two different locations of pristine graphene's
FBZ: $\mathbf{k}=\mathbf{q}$ and $\mathbf{k}=\mathbf{q}+\mathbf{v}_{1}/2$.
Equivalently, we want to construct a basis that verifies $A_{+}\neq0$,
$B_{+}\neq0$ and $A_{-}=B_{-}=0$ when $\mathbf{k}=\mathbf{q}$,
as well as, $A_{+}=B_{+}=0$, $A_{-}\neq0$ and $B_{-}\neq0$ when
$\mathbf{k}=\mathbf{q}+\mathbf{v}_{1}/2$. Given this, we define the
new basis as \begin{subequations} \label{eq:HLenModes} 
\begin{eqnarray}
A_{+} & = & \frac{1}{\sqrt{2}}\big(A_{1}+e^{-iq_{x}a}A_{2}\big),\\
B_{+} & = & \frac{1}{\sqrt{2}}\big(A_{1}+e^{-iq_{x}a}B_{2}\big),\\
A_{-} & = & \frac{1}{\sqrt{2}}\big(A_{1}-e^{-iq_{x}a}A_{2}\big),\\
B_{-} & = & \frac{1}{\sqrt{2}}\big(A_{1}-e^{-iq_{x}a}B_{2}\big).
\end{eqnarray}
 \end{subequations} which verifies the previous conditions. This
new basis uncouples the $(+)$ and the $(-)$ energy sectors of the
doubled unit cell. Note that the new basis written in Eqs. (\ref{eq:HLenModes})
is exactly the same as that defined in Eqs. (\ref{eq:UncpBasis}) and (\ref{eq:zzD-BasisChangeMat}).

\section{The Dirac angle $\theta$}

\label{app:IncdAngle}

\begin{figure}[!ht]
 \centering \includegraphics[clip,width=6cm]{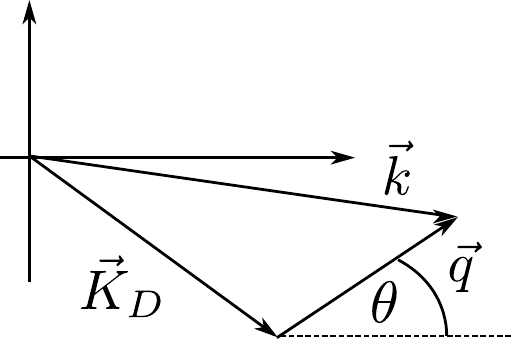}
\caption{(color online) Definition of the Dirac angle $\theta\in[0,\pi]$. }

\label{fig_Dirac_angle_558} 
\end{figure}

In the main text the transmittance is given in terms of $\epsilon$
and $\phi=k_{x} a$. On the other hand, in the figures appearing in the main
text the transmittance is given as function of $\epsilon$ and $\theta$
{[}see. Fig. \ref{fig_Dirac_angle_558} for the definition of $\theta${]}.
We show now how given $\epsilon$ and $\phi$ we can obtain $\theta$.
We consider the case of a doubled unit cell.

For a doubled unit cell Schr\"{o}dinger's equation reads 
\begin{equation}
E\left[\begin{array}{c}
A_{1}\\
B_{1}\\
A_{2}\\
B_{2}
\end{array}\right]=\left[\begin{array}{cccc}
0 & t_{1} & 0 & t_{2}\\
t_{1}^{\ast} & 0 & -te^{i\bm{k}\cdot\bm{u}_{2}} & 0\\
0 & -te^{-i\bm{k}\cdot\bm{u}_{2}} & 0 & t_{1}\\
t_{2}^{\ast} & 0 & t_{1}^{\ast} & 0
\end{array}\right]\left[\begin{array}{c}
A_{1}\\
B_{1}\\
A_{2}\\
B_{2}
\end{array}\right]\,,\label{eq_Hamilt_double}
\end{equation}
 where 
\begin{eqnarray}
t_{1} & = & -t(1+e^{-i\bm{k}\cdot\bm{u}_{2}})\,,\\
t_{2} & = & -te^{-2i\phi}e^{-i\bm{k}\cdot\bm{u}_{2}}
\end{eqnarray}
 Transforming the Hamiltonian to the block diagonal basis we obtain
\begin{equation}
E\left[\begin{array}{c}
A_{+}\\
B_{+}\\
A_{-}\\
B_{-}
\end{array}\right]=\left[\begin{array}{cccc}
0 & t_{+} & 0 & 0\\
t_{+}^{\ast} & 0 & 0 & 0\\
0 & 0 & 0 & t_{-}\\
0 & 0 & t_{-}^{\ast} & 0
\end{array}\right]\left[\begin{array}{c}
A_{+}\\
B_{+}\\
A_{-}\\
B_{-}
\end{array}\right]\,,\label{eq_Hamilt_double_block}
\end{equation}
 where 
\begin{eqnarray}
t_{+} & = & -t(1+e^{-i\bm{k}\cdot\bm{u}_{2}}+e^{-i(\bm{k}\cdot\bm{u}_{2}+\phi)})\,,\\
t_{-} & = & -t(1+e^{-i\bm{k}\cdot\bm{u}_{2}}-e^{-i(\bm{k}\cdot\bm{u}_{2}+\phi)})\,.
\end{eqnarray}
 We can find a set of momentum values where $t_{-}$ is equal to zero.
This corresponds to the new position of the Dirac points in the new
Brillouin zone. The quantity $t_{-}$ can be written as 
\begin{eqnarray}
t_{-} & = & -t(1+e^{-i(\sqrt{3}k_{y}a/2-\phi/2)}-e^{-i(\sqrt{3}k_{y}a/2+\phi/2)})\nonumber \\
 & = & -t\Big(1+2ie^{-i\sqrt{3}k_{y}a/2}\sin\frac{k_{x}a}{2}\Big)\,.
\end{eqnarray}
 If $\sin(k_{x}a/2)=\pm1/2$ and $e^{-i\sqrt{3}k_{y}a/2}=\pm i$ we
have $t_{-}=0$ (and $t_{+}=-2t$). This implies 
\begin{eqnarray}
k_{x}a=\pm\frac{\pi}{3}\,,\\
k_{y}a=\mp\frac{\pi}{\sqrt{3}}\,.
\end{eqnarray}
 Thus, the new Dirac points are located at 
\begin{equation}
\bm{K}_{D}=\left(\pm\frac{\pi}{3a},\mp\frac{\pi}{\sqrt{3}a}\right)\,.
\end{equation}
 The Dirac angle $\theta$ is defined as in Fig. \ref{fig_Dirac_angle_558}.

The dispersion of the low energy modes reads 
\begin{equation}
\epsilon^{2}=\vert t_{-}\vert^{2}/t^{2}=1+4\sin^{2}\frac{\phi}{2}+4\sin\frac{\phi}{2}\sin\frac{k_{y}\sqrt{3}a}{2}\,.
\end{equation}
 The largest and the smallest value of $\phi=ak_{x}$ are given when
$k_{y}=K_{Dy}$. In this case 
\begin{equation}
\epsilon^{2}=(1-2\sin\frac{\phi}{2})^{2}\Rightarrow\sin\frac{\phi}{2}=\frac{1\pm\epsilon}{2}\,.
\end{equation}
 Then 
\begin{equation}
\phi_{min}=2\arcsin\frac{1-\epsilon}{2}\,,
\end{equation}
 and 
\begin{equation}
\phi_{max}=2\arcsin\frac{1+\epsilon}{2}\,.
\end{equation}
 The coordinate $k_{y}a$ is given by 
\begin{equation}
k_{y}a=\frac{2}{\sqrt{3}}\arcsin\frac{\epsilon^{2}-4\sin^{2}\frac{\phi}{2}-1}{4\sin\frac{\phi}{2}}\,.
\end{equation}
 Since $\bm{k}=\bm{K}_{D}+\bm{q}$ we have 
\begin{eqnarray}
q_{x} & = & k_{x}-K_{Dx}=k_{x}-\frac{\pi}{3a}\,,\\
q_{y} & = & k_{y}-K_{Dy}=k_{y}+\frac{\pi}{a\sqrt{3}}\,.
\end{eqnarray}
 From Fig. \ref{fig_Dirac_angle_558} is clear that 
\begin{equation}
\theta=\arccos\frac{q_{x}}{\sqrt{q_{y}^{2}+q_{x}^{2}}}=\arccos\frac{\phi-\pi/3}{\sqrt{(k_{y}a+\pi/\sqrt{3})^{2}+(\phi-\pi/3)^{2}}}\,.
\end{equation}
 Thus, given $\phi$ and $\epsilon$ we can compute $k_{y}a$ and
$\theta$. Scanning $\phi$ between $\phi_{min}$ and $\phi_{max}$
originates $\theta\in[0,\pi]$. Thus $\theta$ is a parametric function
of $\phi$ and $\epsilon$.

\section{The continuum low-energy limit of the TB}

\label{app:CAtransm}

Starting from the results of Ref. \cite{Rodrigues_PRB:2012}, in this appendix we show how can one determine
the general potential governing the behavior of the fermions propagating inside the strip that describes the 
defect line in the continuum low-energy approximation.

As argued in Ref. \cite{Rodrigues_PRB:2012}, in the continuum low-energy limit, one can see the defect line 
as a "strip of width $W$ in the $y$ direction, where there is a general local potential $V(y) = V_{s} 
\mathbb{I} + V_{x} \sigma_{x} +V_{y} \sigma_{y} + V_{z} \sigma_{z}$ for $\vert y \vert < W/2$". Therefore, in 
that region, the Dirac Hamiltonian reads
\begin{eqnarray}
  H &=& \hbar v_{F} \big(\nu\sigma_{x}(-i\partial_{x})+\sigma_{y}(-i\partial_{y})\big)+\big(V_{s} \mathbb{I}
  +\mathbf{V}\cdot \boldsymbol{\mathbf{\sigma}}\big) . \label{eq:DiracHamPot}
\end{eqnarray}

Taking the continuum low-energy limit of such a system, is equivalent, either to take the limit 
$W\times(V_{s},\mathbf{V})\to(v_{s},\mathbf{v})$ when $W\to0$ keeping $\epsilon$ and $q_{x}$ finite, 
or to take the limit $\epsilon,q_{x}\to0$ keeping $W$ finite.

In either case, one can write a boundary condition matrix, $\mathcal{M}$, relating the wave-function 
at each side of the defect line, $\Psi(x,W/2)=\mathcal{M}\Psi(x,-W/2)$. In general it reads 
\cite{Rodrigues_PRB:2012}
\begin{equation}
  \mathcal{M} = e^{-i\sigma_{y}(v_{s} \mathbb{I} + \mathbf{v} \cdot \boldsymbol{\sigma})/v_{F}} , \label{eq:BCgeneral}
\end{equation}
where $\boldsymbol{\sigma}=(\sigma_{x},\sigma_{y},\sigma_{z})$ stand for the Pauli matrices, while $v_{F}$ is
the Fermi velocity.

Defining $\alpha := \sqrt{v_{s}^{2} - (v_{x}^{2}+v_{z}^{2})}/v_{F} \hbar$, and using $\sigma_{i} \sigma_{j} = 
\delta_{i j} i \epsilon_{i j k} \sigma_{k}$, it is simple to show that
\begin{eqnarray}
  \mathcal{M} &=& e^{-i v_{y}/\hbar v_{F}} \left ( \begin{array}{cc} \cos \alpha - \frac{v_{x}}{\hbar v_{F} 
  \alpha} \sin \alpha & \frac{v_{z} - v_{s}}{\hbar v_{F} \alpha} \sin \alpha \\ \frac{v_{z} + v_{s}}{\hbar v_{F} 
  \alpha} \sin \alpha & \cos \alpha + \frac{v_{x}}{\hbar v_{F} \alpha} \sin \alpha \end{array} \right ) . 
  \label{eq:MMalpha}
\end{eqnarray}

For the {\it pentagon-only} defect line, the continuum low-energy boundary condition matrix, $\mathcal{M}_{5 5}$,
reads \cite{Rodrigues_PRB:2012}
\begin{eqnarray}
  \mathcal{M}_{5 5} &=& \left ( \begin{array}{cc} 0 & 1 \\ -1 & \xi \end{array} \right ) . \label{eq:Mca55}
\end{eqnarray}

The potential terms $(V_{s},\mathbf{V})$ thus read
\begin{subequations} \label{eq:GenPotential}
\begin{eqnarray}
  V_{s} &=& \frac{v_{s}}{W} = - \frac{\hbar v_{F} \alpha}{W \sin \alpha} , \\
  V_{x} &=& \frac{v_{x}}{W} = \frac{\cos \alpha}{W \sin \alpha} \hbar v_{F} \alpha , \\
  V_{z} &=& \frac{v_{z}}{W} = 0 ,
\end{eqnarray}
\end{subequations}
while $V_{y}$ is arbitrary, and thus can be taken equal to $0$. Moreover, we can do the identification
\begin{eqnarray}
2 \cos \alpha &=& \xi .
\end{eqnarray}

The term $V_{s}$, is a mass term, that only causes a shift in the energy of the massless Dirac fermion. This
term will deviate the fermion's direction of propagation inside the strip when compared to its direction outside
the strip. The term $V_{x}$ deviates the fermion's direction of propagation and tilts its spin. The 
spin of the fermion will no longer be aligned along its direction of propagation as happens for fermions 
outside the strip.

For the sake of comparison, let one write the eigenstate associated with a free massless Dirac fermion propagating 
outside the strip [Eq. (\ref{eq:DiracHamPot}) with $(V_{s},\mathbf{V})=(0,\mathbf{0})]$ and that of a massless 
fermion propagating inside the strip [Eq. (\ref{eq:DiracHamPot}) with $V_{y} = 0 = V_{z}$]. The first can be written 
as
\begin{eqnarray}
  \big\vert \psi_{\nu}^{\gtrless} \big\rangle &=& \frac{1}{\sqrt{2}} \left ( \begin{array}{c} 1 \\ 
  s \nu e^{\pm i \nu \theta} \end{array} \right ) e^{i (q_{x} x \pm q_{y} y)}, \label{eq:3}
\end{eqnarray}
while the second reads
\begin{eqnarray}
  \big\vert \phi_{\nu}^{\gtrless} \big\rangle &=& \frac{1}{\sqrt{2}} \left ( \begin{array}{c} 1 \\ 
  \widetilde{s} e^{\pm i \beta_{\nu}} \end{array} \right ) e^{i (\widetilde{q}_{x} x \pm \widetilde{q}_{y} y)}. 
  \label{eq:4}
\end{eqnarray}
In the above expressions, the symbol $>$ ($<$) identifies the state propagating to $y\to+\infty$ ($y\to-\infty$),
and $\nu=\pm1$ identifies the Dirac point $\mathbf{K}_{\nu}$ the eigenstate refers to. The symbol 
$s$ identifies the sign of the energy of the free fermion, $s=\textrm{Sign}[\epsilon]$, while the angle 
$\theta = \arctan(q_{y}/q_{x})$ determines the orientation of its spin. Similarly, $\widetilde{s} 
= \textrm{Sign} [\epsilon-v_{s}]$ stands for the sign of the energy of the fermion propagating inside the strip, 
while 
\begin{eqnarray}
  \beta_{\nu} &=& \arctan \Bigg(\frac{\widetilde{q}_{y}}{\nu \widetilde{q}_{x}+v_{x} /W v_{F} \hbar}\Bigg) ,
\end{eqnarray}
determines the orientation of its spin.

The translation symmetry along the defect line, forces $\widetilde{q}_{x} \equiv q_{x} = \vert \mathbf{q} \vert \cos 
\theta = ( \epsilon / v_{F} \hbar) \cos \theta$. Then, $\widetilde{q}_{y}$ can be determined from the energy, 
$\epsilon$, and the angle of incidence in the strip, $\theta$. It reads
\begin{eqnarray}
  \widetilde{q}_{y} &=& \sqrt{\bigg(\frac{\epsilon - v_{s}/W}{v_{F} \hbar}\bigg)^{2} - \bigg(\frac{\nu \epsilon \cos 
  \theta + v_{x}/W}{v_{F} \hbar}\bigg)^{2}}. \label{eq:qytildeStrip}
\end{eqnarray}

Very near the Dirac points, $\epsilon, q_{x} \to 0$, all the Dirac fermions travelling inside the strip
will have the same $\beta_{\nu}$
\begin{eqnarray}
  \beta_{\nu} &=& \arctan \Bigg(\frac{\sqrt{v_{s}^{2}-v_{x}^{2}}}{v_{x}}\Bigg) . \label{eq:alphaZero}
\end{eqnarray}
From Eqs. (\ref{eq:GenPotential}) and Eq. (\ref{eq:alphaZero}), one can easily conclude that when
$\epsilon, q_{x} \to 0$, we have that $\alpha \equiv \beta_{\nu}$. Thus, in the low-energy limit, all the 
fermions propagating inside the strip, will have their spins aligned in the same direction. As discussed
in the main text, perfect transmission occurs when $\theta = \beta_{\nu}$.

In a similar way, we can also compute the general potential describing both the $zz(558)$ and the $zz(5757)$
defect line in the continuum low-energy limit. The boundary condition matrix {\it seen} by the massless Dirac 
fermions at the $zz(558)$ and at the $zz(5757)$ defect lines are computed in Ref. \cite{Rodrigues_PRB:2012}.
The one originating from the $zz(558)$ defect line reads
\begin{eqnarray}
  \mathcal{M}_{5 5 8} &=& \left ( \begin{array}{cc} 0 & 1 \\ -1 & 2 \frac{\xi_{2}}{\xi_{1}^{2}} \end{array} 
  \right ) . \label{eq:Mca558}
\end{eqnarray}
This boundary condition is very similar to the one computed for the {\it pentagon-only} defect line, Eq. 
(\ref{eq:Mca55}). They are equal if we define an effective hopping parameter $\widetilde{\xi} := 2 
\xi_{2}/ \xi_{1}^{2}$. Therefore, the corresponding general potential is going to be given by 
\begin{subequations} \label{eq:GenPotential-558}
\begin{eqnarray}
  V_{s}^{558} &=& \frac{v_{s}^{558}}{W} = - \frac{\hbar v_{F} \widetilde{\alpha}}{W \sin \widetilde{\alpha}} , \\
  V_{x}^{558} &=& \frac{v_{x}^{558}}{W} = \frac{\cos \widetilde{\alpha}}{W \sin \widetilde{\alpha}} \hbar v_{F} 
  \widetilde{\alpha} , \\
  V_{y}^{558} &=& \frac{v_{y}^{558}}{W} = 0 = \frac{v_{z}^{558}}{W} =  V_{z}^{558} ,
\end{eqnarray}
\end{subequations}
where $\widetilde{\alpha} = \arccos (\xi_{2}/ \xi_{1}^{2})$ is the angle the spin of the fermions 
propagating inside the strip will make with the horizontal direction, in the limit where $\epsilon$, $\phi \to 0$.

From the boundary condition matrix originating from the $zz(5757)$ defect line,
\begin{eqnarray}
  \mathcal{M}_{5757}&=& \left(\begin{array}{cc}
      -a/d & -b/d \\ b/d & -c/d \end{array}\right),\label{eq:Mca5757}
\end{eqnarray}
where $a = 2 \xi_{c}^{2}\big(\xi_{b}^{2}-\xi_{a}^{2}/4\big)$, $b=-\xi_{a}(\xi_{b}^{2}-\xi_{a}^{2})$,
 $c=2(\xi_{b}^{4}+\xi_{a}^{4}+\xi_{b}^{2}\xi_{a}^{2})/\xi_{c}^{2}$ and $d = 2 \xi_{b}(\xi_{b}^{2}+\xi_{a}^{2}/2)$,
one can write the following general potential 
\begin{subequations} \label{eq:GenPotential-5757}
\begin{eqnarray}
  V_{s}^{5757} &=& \frac{v_{s}^{5757}}{W} = - \frac{\hbar v_{F} \bar{\alpha}}{W \sin \bar{\alpha}} \frac{b}{d} , \\
  V_{x}^{5757} &=& \frac{v_{x}^{5757}}{W} = \frac{\cos \bar{\alpha} + a/d}{W \sin \bar{\alpha}} \hbar v_{F} 
  \bar{\alpha} , \\
  V_{y}^{5757} &=& \frac{v_{y}^{5757}}{W} = 0 = \frac{v_{z}^{5757}}{W} = V_{z}^{5757} ,
\end{eqnarray}
\end{subequations}
where $\bar{\alpha} = \arccos [ -(a+c)/ 2 d]$ is the angle the spin of the fermions 
propagating inside the strip will make with the horizontal direction, in the limit where $\epsilon$, $\phi \to 0$.

\begin{ack} J. N. B. R. was supported by Funda\c{c}\~{a}o para a Ci\^{e}ncia
e a Tecnologia (FCT) through Grant No. SFRH/BD/44456/2008. N. M. R.
P. was supported by Fundos FEDER through the Programa Operacional
Factores de Competitividade - COMPETE and by FCT under project no.
PEst-C/FIS/UI0607/2011. N. M. R. P. acknowledges both the hospitality and the 
funding from the Graphene Research Centre at the National University of 
Singapore, where this work was completed.

\end{ack}


\section*{References}
\providecommand{\newblock}{}


\end{document}